\documentclass[10pt,preprint]{aastex}
\input epsf.tex
\begin{document}
\title{The White Dwarf Cooling Sequence of NGC~6397\altaffilmark{1}}
\author{Brad M. S. Hansen\altaffilmark{2,3}, Jay Anderson\altaffilmark{4},
James Brewer\altaffilmark{5}, Aaron Dotter\altaffilmark{6},
 Greg. G. Fahlman\altaffilmark{5,7}, Jarrod
Hurley\altaffilmark{8}, Jason Kalirai\altaffilmark{9}, Ivan King\altaffilmark{10},
David Reitzel\altaffilmark{2}, Harvey B. Richer\altaffilmark{5}, R.Michael Rich\altaffilmark{2}, Michael M. Shara\altaffilmark{11}, Peter B. Stetson\altaffilmark{12}}
\altaffiltext{1} {Based on observations with the NASA/ESA Hubble Space Telescope, obtained at the Space Telescope Science
Institute, which is operated by the Association of Universities for Research in Astronomy, Inc., under NASA contract NAS 5-26555.
These observations are associated with proposal GO-10424}
\altaffiltext{2}{Department of Astronomy, University of California Los Angeles, Los Angeles, CA 90095, hansen@astro.ucla.edu, reitzel@astro.ucla.edu, rmr@astro.ucla.edu}
\altaffiltext{3}{Institute for Geophysics and Planetary Physics,University of California Los Angeles, Los Angeles, CA 90095}
\altaffiltext{4}{Department of Physics and Astronomy, Rice University, Houston, TX 77005, jay@eeyore.rice.edu}
\altaffiltext{5}{ Department of Physics \& Astronomy, 6224 Agricultural Road, University of British Columbia, Vancouver, BC, V6T 1Z4, Canada,  richer@astro.ubc.ca }
\altaffiltext{6}{Department of Physics \& Astronomy, Dartmouth College, 6127 Wilder Laboratory, Hanover, NH, aaron.l.dotter@dartmouth.edu}
\altaffiltext{7}{ Canada-France-Hawaii Telescope Corporation, P.O. Box 1597, Kamuela, HA, 96743, fahlman@cfht.hawaii.edu}
\altaffiltext{8}{Department of Mathematics and Statistics, Monash University, Clayton, Victoria 3800, Australia, jarrod.hurley@sci.monash.edu.au}
\altaffiltext{9}{Hubble Fellow, Lick Observatory,  University of California Santa Cruz, Santa Cruz, CA, 95064, jkalirai@ucolick.org}
\altaffiltext{10}{Department of Astronomy, University of Washington, Seattle, WA 98195, king@astro.washington.edu}
\altaffiltext{11}{ Department of Astrophysics, Division of Physical Sciences, American Museum of Natural History, Central Park West at
79th St, New York, NY, 10024-5192, mshara@amnh.org}
\altaffiltext{11}{Hertzberg Institute of Astrophysics, National Research Council, Victoria, BC, Canada, peter.stetson@nrc-cnrc.gc.ca}

\slugcomment{\it 
}

\lefthead{Hansen {\it et al.}}
\righthead{NGC6397 White Dwarfs}

\begin{abstract}

We present the results of a deep Hubble Space Telescope (HST) exposure of the nearby globular cluster NGC~6397, focussing attention
on the cluster's white dwarf cooling sequence. This sequence is shown to extend over 5 magnitudes in depth,
with an apparent cutoff at magnitude $F814W \sim 27.6$. We demonstrate, using both artificial star tests
and the detectability of background galaxies at fainter magnitudes, that the cutoff is real and represents
the truncation of the white dwarf luminosity function in this cluster. We perform a detailed comparison
between cooling models and the observed distribution of white dwarfs in colour and magnitude, taking into
account uncertainties in distance, extinction, white dwarf mass, progenitor lifetimes, binarity and cooling 
model uncertainties. After marginalising over these variables, we obtain values for the cluster distance
modulus and age of $\mu_0 = 12.02 \pm 0.06$ and $T_c = 11.47 \pm 0.47$~Gyr (95\% confidence limits).
Our inferred distance and white dwarf initial-final mass relations are in good agreement with other independent
determinations, and the cluster age is consistent with, but more precise than, prior determinations made
using the main sequence turnoff method. In particular, within the context of the currently accepted
$\Lambda$CDM cosmological
model, this age places the formation of NGC~6397 at a redshift $z \sim 3$, at a time when 
the cosmological
star formation rate was approaching its peak.

\end{abstract}


\section{Introduction}


The oldest known stellar systems in the Milky Way are the globular clusters. As such, their
nature reflects the conditions under which our Galaxy first formed and offers a unique window
into the high redshift universe. Age determination of globular clusters
by fitting models to the main sequence turnoff has a long and venerable history (Sandage 1953; Janes \& Demarque 1983; Fahlman, Richer \& VandenBerg 1985; Chieffi \& Straniero 1989; VandenBerg, Bolte \& Stetson 1996 and references therein). 
For a long period of time, the results of these studies led to a so-called `cosmological crisis',
in which the estimated cluster ages ($\sim 16-20$~Gyr) were larger than the age of the
universe based on a variety of cosmological tests (e.g. Bolte \& Hogan 1995). However, over the last decade, improvements
in the distance measurements to globular clusters, particularly using the Hipparcos satellite,
 have resulted in a lower estimate for the mean age of the metal poor (hence oldest) globular clusters
($>10.4$~Gyr at $95\%$ confidence; Krauss \& Chaboyer 2003), which is now consistent with the age of the universe estimated from microwave
background measurements (Spergel et al 2003).  
As befits a mature method, the accuracy of this MSTO method is currently limited
by a variety of systematic errors (distance and metallicity uncertainties, atmosphere models that
don't fit the shape of the turnoff). This method has been carried as far as it can go with current technology
and further significant improvements must await technical advances, such as improvements in distance measurement
using the 
Space Interferometry Mission (Chaboyer et al 2002).

In the past several years, we have embarked on a program to measure the ages of  globular
clusters by an entirely different method -- measuring the white dwarf cooling sequence (WDCS)
and determining the age by modeling the rate at which they cool (Hansen et al 2002; 2004).
This method also has a distinguished history when applied to the stellar population in the
solar neighbourhood (Winget et al 1987; Wood 1992; Hernanz et al 1994; Oswalt et al 1996; Leggett et al. 1998; Hansen 1999; Harris et al 2006) but it has only recently become possible
to apply the same method to globular clusters because of the extreme requirements such
measurements place on both
resolution and photometric depth.

Our initial measurements of the age of Messier~4~(M4), the closest globular cluster to the sun, yielded
a best-fit age of 12.1~Gyr, with a 95\% lower bound of 10.3~Gyr. This is similar to the accuracy
achieved by the latest MSTO analysis ($12.6^{+3.4}_{-2.2}$~Gyr) by Krauss \& Chaboyer (2003).
The M4 measurement was performed using the WFPC2 camera on the Hubble Space Telescope,
which has since been superceded by the Advanced Camera for Surveys (ACS). The larger field
and better resolution of ACS thus offers the possibility of improving on this result.
Furthermore, the original M4 result was considered controversial by some (de Marchi et al. 2004), 
although unreasonably so in our opinion (Richer et al 2004; Hansen et al 2004). Nevertheless,
an opportunity to repeat the experiment would serve to dispel any lingering doubts.

Consequently, we have observed a field in the second closest globular cluster to the sun, NGC~6397,
using ACS on HST, with the goal of measuring the white dwarf cooling sequence to unprecedented
depth and precision. In addition, the lower reddening (relative to M4) of this cluster and the
inclusion of a series of short exposures (to avoid saturation of bright stars), allows us to
perform a fit to the MSTO (Richer et al. 2007) simultaneously while fitting the WDCS -- a measurement never before
possible.  In \S~\ref{WDobs}, we describe the observations and the reduction process, including characterisation and removal of background galaxies as well as the characterisation of our observational
biases, using extensive artificial star tests. In \S~\ref{WDmodels}, we describe our modeling procedure and construct simulated cooling sequences which we then compare to the data. In \S~\ref{WDDiscuss} we summarise our result and its implications.

\section{Observations}
\label{WDobs}
The observational data consist of a series of images taken with 
ACS of a single
field in NGC~6397, over the course of 126 orbits (4.7 days). The field
is located 5 arcminutes southeast of the cluster core, overlapping that
of several previous data sets taken by the Wide~Field~Planetary~Camera~2 (WFPC2) in 1994 \& 1997.
 All told the
data set consists of 252 exposures (totalling 179.7 ksec) in the F814W filter
and 126 exposures totalling 93.4 ksec in F606W. The F814W exposures were taken
at the beginning and end of each visibility period, so that the F606W exposures were taken
with as low a sky background as was possible. The images were typically dithered by 15 pixels, with additional sub-pixel dithers applied. Apart from a variety
of short exposures (durations ranging from 1s to 40s) used to treat the brighter stars, the deep exposures were of similar duration and were thus assigned equal weight in the analysis. The full details of the analysis will be described
elsewhere (Anderson et al 2007) but we review here the aspects thereof that
are directly relevant to the identification of the faint point sources that make up the white dwarf cooling sequence.

    Essentially, our approach uses the fact that the only 
    influence the faintest stars will have on the image is to push their 
    central pixels up above the noise in some number of the images.  If a star is 
    not bright enough to do this, then there is no way it will be found.  
    However, if it does manage to generate a local maximum in a 
    statistically significant number of images, then one can find it by 
    searching for coincident peaks, peaks that occur in the same place 
    in the field in different exposures.  After much experimentation 
    (including artificial star tests), it was determined that a star must 
    generate a peak in at least 90 out of the 252 F814W images to qualify 
    as a 99\% significance detection (the inclusion of the F606W images 
    did not increase the significance levels).

    Thus, we generate a list of possible sources by identifying every place
    in the field where a peak was found in at least 90 F814W images.  To
    suppress artifacts around bright stars, we also require that the source 
    be the most significant concentration of peaks within 7.5 pixels.  This
    initial pass results in a catalog of 48,785 potential sources.  
    Inspection of these sources on a stacked image shows that while this procedure 
    identifies everything that is visibly a star, it also
    nets plenty of artifacts and galaxies.  

    We measure each of these sources as if it was a star, using a careful 
    PSF-fitting procedure to get a position and a flux in F606W and F814W.
    We then use a three-pronged weeding procedure to remove the non-stellar 
    objects.  The first two procedures target the PSF artifacts.  While the 
    7.5-pixel halo removes a lot of such artifacts, there are many PSF features 
    that extend beyond this radius.  A spike filter is applied to the stacked 
    image to identify the likely contribution of diffraction spikes to every 
    point in the field.  We require that each identified source be much brighter than the
    expectation value of spikes at its location.  The second weeding step
    examines the brightness of a source in relation to its brighter neighbors.
    Artifacts exhibit a clear relationship between their distance from a bright
    star and their own brightness.  We create a halo around each bright star
    that tells us how bright a faint star must be to be believable.  Both of
    these tests are detailed in Anderson et al (2007).

    The above cuts remove the instrumental artifacts but still include real astrophysical objects (i.e. galaxies)
 that we wish to discount. In order to restrict our sample to point sources we make two further morphological cuts.

\subsection{Star-Galaxy Separation}

The first cut compares the flux in the central pixel to the overall flux
extracted from the PSF model for the star. From this we measure the fractional
central pixel excess/deficit, hereafter termed {\tt CENXS}. A deficit indicates that the object is partially
resolved and thus extended. A similar measure ({\tt ELONG})  can be made for the residual asymmetry in the image within 1.5 pixels radius. Larger asymmetries again suggest that the object is resolved. On this basis we can remove all the partially resolved
 objects, leaving us with a sample of 8399 potential point sources.

The last cut made is to exclude a region of radius 180~pixels around an apparent
concentration of faint blue sources located at (x,y)=(725,720) on the chip. It
turns out that there is a 
 large
elliptical galaxy located near that position.
Thus these objects are most likely not white dwarfs
but rather globular clusters associated with the galaxy. Most of these
fall outside the colour-magnitude range of the white dwarfs, but not all of
them, and so we exclude sources in this region from our sample.

\subsection{Magnitude Calibration}

The conversion from instrumental to Vega magnitudes was done according to the
prescription of Sirianni et al. (2005). We measure the bright stars on specific
reference images in F814W ({\tt j970101bbq\_drz}) and F606W ({\tt j970101bdq\_drz}), using
the standard procedure and 10-pixel radius aperture. Because this is a crowded field, we 
reject some of these stars for which too large a fraction of the flux was found beyond 5 pixels.
This controls errors that might be introduced by either cosmic rays or other stars. 
These are then compared to the fluxes measured using our PSF-fitting procedure to determined
the zero-point for our calibration. The sigma-clipped average of this comparison yields
 ZP(F606W)=33.321 and ZP(F814W)=32.414.

Figure~\ref{pretty_cmd} shows the final colour-magnitude diagram that results.
The main sequence of the cluster is clearly seen, from above the turnoff to
where the contrast with the background population fades near the hydrogen burning limit. Most exciting for the purposes of this paper, is the very clear cooling sequence of the cluster white dwarfs, beginning near F814W$ \sim 22.5$ and extending down to an observed truncation at F814W$ \sim 27.6$. The presence of a population
of bluer and fainter objects (mostly the remaining unresolved galaxies but perhaps also a few
background white dwarfs) in the CMD  indicates that the truncation is real, and
not a result of observational incompleteness.
 Figure~\ref{pretty_wd}
shows  a zoom into the faint, blue region of this diagram -- where white dwarfs and galaxies
lie. The two panels show the effect of our  galaxy--star separation.
Due to the excellent image quality of the ACS data, we can
identify most of the galaxies from the photometry alone. While there is certainly
still some level of galaxy contamination in the point source sample, it is now
at a level that can be securely modeled and included in the errors. As we will
show in subsequent sections, with this data set it is possible to do model comparisons
even without a proper motion separation.

\subsection{Artificial Stars}


An important part of the modeling effort is the need to subject the models to the same
kind of observational scatter and incompleteness as the data. This is particularly important
for a data set such as this, where even relatively bright white dwarfs can be lost if they
happen to be projected close to a bright main sequence star. Thus, we have performed a
detailed series of artificial star tests, to measure not only the recovery fraction, but also
the degree of correlation between input magnitude and observed magnitude.

Artificial stars were chosen with magnitudes and colours drawn from the canonical white dwarf
cooling sequence, and added to the data on a grid, with mutual separation of ten pixels, plus
a random offset of 0.5 pixels relative to the grid position. This
is large enough that any given star does not affect the detectability of its neighbours. The
new data set is then analysed as before and the recovery fraction measured. Figure~\ref{Rec} shows
the resulting recovery fraction as a function of F814W. Also shown are the measured dispersions
of the recovered magnitudes as a function of input magnitude. We see that, down to magnitude
F814W=28, the recovery fraction is $>50\%$ and the scatter is $<0.25$ mags.

\subsection{Residual Galaxy Contamination}

In order to model the distribution of white dwarfs in colour and magnitude,
we bin the data in the manner shown in Figure~\ref{WDgrid}. As described in
Hansen et al (2004) we find that the distribution of white dwarfs above the
luminosity function jump (F814W$\sim 26$ in this case) contain very little
age information, so these are added together into a single large bin. However,
at fainter magnitudes, the distribution in magnitude and in
colour is affected by the mass and age of the underlying white dwarfs and
so 
 there is useful information to be obtained by quantifying this distribution.
Thus we bin our data on a grid in colour--magnitude space from 
F606W-F814W$=0.9$ to 1.5 and from F814W=26 to 28.

Our definition of our data product thus far involves the cut on the {\tt CENXS} and 
{\tt ELONG}  parameters
to exclude extended sources, followed by binning the data as described. Before we
continue to the modeling we also require a measure of the systematic error contributions
to the counts in each bin. The first such contribution is from the 
 remaining galaxy contamination
in each bin. Figure~\ref{SharpDis} shows the distribution of {\tt CENXS } parameter for {\em all} 
objects in the range $26<$F814W$<28$ and $0.9<$F606W-F814W$<1.5$. Hence this data set represents
the distribution of all objects that fall within the colour-magnitude bounds of interest
in describing the white dwarf cooling sequence. We model this as a sum of two gaussians.
One, the narrow peak, represents the stellar component and the second represents the broad
distribution of {\tt CENXS} parameters appropriate to the partially resolved galaxy contribution.
We will use this second distribution to estimate the potential contribution of unresolved
galaxies to our model fits. A second contribution to systematic error comes from the fact that
the counts will differ slightly if we change the particular values of the cuts on {\tt CENXS}
and {\tt ELONG}, so we include a contribution in each bin due to a reasonable variation in
the actual values of the cuts.


Table~\ref{CMDtable} shows the number counts of sources with $\left| \right.${\tt CENXS}$ \left. \right|<0.02$ 
in each bin along with the estimated error in each bin, including both statistical and
systematic error.



\subsection{An empirical cooling sequence}

Before we proceed to the full modeling description, we pause to describe a procedure whereby
we can derive an empirical cooling sequence, without reference to any modeling. Armed with
the data and the artificial star tests, we can derive an empirical relationship between 
colour and magnitude under the simple assumption that the underlying relationship is monotonic
-- that, for any given magnitude, there is a unique colour appropriate to the underlying cooling
sequence. The observed scatter is assumed to be the result of photometric errors, quantified by
the artificial star tests.

Our derivation, briefly described in Richer et al (2006), proceeds as follows. For each magnitude bin defined in Figure~\ref{WDgrid}, our
data consists of number counts as a function of colour. If we assume a value for the intrinsic
model colour at this magnitude, we can then use the results of 
 the artificial star tests to predict the distribution of colour (at this magnitude)
 expected after accounting for observational scatter. We may then characterise
how well this fits with the true observed colour distribution using the $\chi^2$ statistic. 
By varying the intrinsic model colour until we find the minimum of $\chi^2$ at each magnitude,
we may then derive the best fit colour as a function of magnitude -- an empirical cooling sequence.

This is shown in Figure~\ref{Fit}. This may seem superfluous given that, in subsequent
sections, we're going to fit model atmosphere colours to the data. However, there are still 
several issues outstanding in the chemical evolution and atmospheric modeling of white dwarfs 
which means that there is some uncertainty in the final model colours (Bergeron, Ruiz \& Leggett 1997;
Bergeron \& Leggett 2002). So, it is of interest to see what kind of relationship between
colour and magnitude fits the data independent of theoretical models. In particular we see evidence
in Figure~\ref{Fit}
(in the form of a turn towards the blue)
for the deviation from black body trends (Hansen 1998; Saumon \& Jacobsen 1999) expected due to collisionally induced absorption by
molecular hydrogen in hydrogen-rich white dwarf atmospheres (Mould \& Liebert 1978; Bergeron, Saumon \& Wesemael 1995; Borysow, Jorgensen \& Zheng 1997). 
Below we will see that this empirical relationship is similar to that found from theoretical hydrogen atmosphere models,
suggesting that our sample is dominated by Hydrogen atmosphere dwarfs.

Furthermore, this empirical sequence should allow other groups to compare their models to the
data. Table~\ref{EmpTab} gives the best fit colours as a function of magnitude.

\subsection{Distance and Extinction}

In both the MSTO and WDCS methods, the determination of the distance to the cluster is a fundamental aspect of the age measurement.
 The traditional method for globular clusters is to compare the 
main sequence with local, metal-poor subdwarfs with known parallaxes to determine the
distance. We take our default distance to NGC~6397 to be
$\mu_0 = 12.13 \pm 0.15$ by Reid \& Gizis (1998), who used 
 Hipparcos distances for the subdwarfs in V \& I for their main sequence fit. This also assumes a reddening E(B-V)=0.18
for this line of sight. We chose this determination because most other main sequence distance determinations to NGC~6397 use the B \& V bandpasses, so that the Reid \& Gizis work is a more direct comparison to the bandpasses used here. We will examine this further in \S~\ref{Distance}.

Although we shall later compare to the distance and extinction derived from the main sequence,
we prefer to initially determine these quantities directly from the white dwarf sequence. There
are several reasons for this preference -- for a cluster as metal poor as NGC~6397, there
are very few appropriate subdwarfs and colour transformations are necessary; our observations
use the F606W bandpass rather than the F555W, so that there are non-negligible colour transformations
between the HST bandpasses and the ground-based bandpasses used for the distance and extinction
transformation, and finally there is the fact that cool white dwarfs can have spectral shapes
not well represented by the (usually much hotter) stars used to determine the reddening.

To determine the distance and extinction, we follow the same method as used in Hansen et al (2004),
essentially a variation of the distance determination method of Renzini et al (1996).
An empirical fit to the upper part of the observed NGC~6397 cooling sequence is
\begin{equation}
{\rm F814W} = 3.00 \left( {\rm F606W-F814W} \right) +23.37  \label{EmpFit}
\end{equation}
Using the atmosphere models of Bergeron et al (1995), but corrected slightly to be at fixed radius rather than
fixed gravity,  
 we get
\begin{equation}
{\rm (F814W)_0}= 2.77 \left({\rm  F606W-F814W} \right)_0 + 11.51 - 5 \log R_9   \label{BFit}
\end{equation}
where $R_9$ is the white dwarf radius in units of $10^9$cm and the subscript $0$ indicates the
colour before reddening is applied. Taking reddening into account and assuming
$A_{814}=0.65 A_{606}$ (Sirianni et al 2005), we can express the true distance modulus in terms of the other
fitting parameters
\begin{equation}
\mu_0 = 11.86 + 0.49 A_{814} + 5 \log R_9 + 0.23 \left({\rm F606W-F814W}\right)
\end{equation}

The last term indicates that there is some small colour dependance in the fit (resulting from the fact that
there is a residual difference in the slopes between equations~\ref{EmpFit} and \ref{BFit}). The other
two parameters are the extinction and the white dwarf radius (related directly to the mass).
Thus, one can trade off the distance, extinction and white dwarf mass to maintain a
good fit to the observed cooling sequence. This means that the constraints on these three
quantities are interrelated. Using an extinction of $A_{814}=0.3$ (explained below), and 
performing the fit at F606W-F814W$=0.6$, we find the distance and radius are related by
\begin{equation}
\mu_0 = 12.15 + 5 \log R_9. \label{muR1}
\end{equation}

For a 0.5$M_{\odot}$ white dwarf at the top of the cooling sequence, we get $\mu_0=12.08$, which is well
within the range of acceptable 
 distance moduli derived by Reid \& Gizis. If we consider the lower limit from Reid \& Gizis to be
$\mu_0>12.0$, this restricts the range of allowed radii and hence places an upper limit on the masses at
the top of the cooling sequence $M < 0.53 M_{\odot}$.

Of course, this just means that the models and observations agree at the bright end. Much
more information emerges when we consider the data at the fainter end. In the following
sections, we will get constraints on the distance, extinction and masses directly from
the fit to the full cooling sequence. But we can get a preview of these results in a simple way 
 by performing an operation similar to the one above, but now for the fainter white dwarfs.
In this case, the colour shift at faint magnitudes provides a feature in the cooling curve
which must be fit by the observations. The most transparent way to do this is to
compare our empirical cooling curve with
the model curve of colour versus magnitude. Fitting a $0.5 M_{\odot}$ model curve to the empirical
sequence (Figure~\ref{Fit}) suggests a reddening in the observed passbands of E(F606W-F814W)$=0.16$. If
we infer from this an extinction $A_{814}=0.30$, then fitting a 0.5$M_{\odot}$ white dwarf model
to the data requires 
$ \mu_0 = 12.0$,
which is slightly smaller than the value 
derived in equation~\ref{muR1}. 
The two independent measures can be made consistent if we allow the mass to increase along the
cooling sequence, since the difference in the inferred $\mu_0$ is then offset by the more negative
value of the
 $5 \log R_9$ term for a larger mass white dwarf.
 By requiring that our model fit both constraints simultaneously,
we are thus able to place limits on the initial-final mass relation.
The fact that the change is not large also suggests that there is not a lot of variation in mass between
the bottom and top of the cooling sequence, as larger mass models have smaller radii and would result
in a smaller inferred $\mu_0$. 

We will
 examine all of this in more detail in \S~\ref{WDmodels}, but the above analysis demonstrates
the origin of some of our forthcoming constraints on cluster distance and 
white dwarf masses. It also suggests our measurements are not particularly sensitive
to many of the details of the initial-to-final mass relation (which emerge mostly at the high-mass end).
This will also be examined in \S~\ref{IFMRsec}.

\section{Models}
\label{WDmodels}

The final data product is shown in Figures~\ref{LF0} and \ref{Hess0}. In Figure~\ref{LF0} we
show the white dwarf luminosity function using a cut $\left| \right. ${\tt CENXS}$ \left. \right| < 0.02$
and {\tt ELONG}$<0.02$. Figure~\ref{Hess0} shows the binned data from the Hess diagram
 in colour-magnitude space using the same cuts. In Figure~\ref{LF0} we also show the galaxy luminosity
function, which rises smoothly well beyond the observed truncation in the white dwarf luminosity function.

\subsection{Default Models}
\label{Default}

We are now in a position to begin our investigation of fitting models to the
full cooling sequence. Our initial fit will be to the same default
model as in Hansen et al (2004).
We use the white dwarf cooling models of Hansen (1999), with a hydrogen layer mass fraction of
$10^{-4}$ and a helium layer mass fraction of $10^{-2}$.
 We furthermore adopt an initial--final mass relation
of the form ($M_{TO}$ is the turnoff mass at a given age)
\begin{equation}
M_{wd} = A e^{B (M-M_{TO})} \label{Mwdfit}
\end{equation}
and we determine $A$ and $B$, along with cluster distance $\mu_0$ and extinction $A_{814}$, by
minimizing the $\chi^2$ of the model fit to the observations, in tandem with the determination
of cluster age and progenitor mass function slope. To begin with, we take the lifetimes of main sequence
progenitors from the same models as in Hansen et al (2004).
We will investigate the effect of other main sequence models
in \S~\ref{LowZ}.

Using this model we generate a Monte Carlo realisation of the white dwarf population.
For each white dwarf, we then add a photometric error 
chosen from the
distribution of output magnitudes (given the intrinsic value as input magnitude) as
determined from the artificial star tests. We do this for both F606W and F814W.
 In this manner, each realisation includes
a realistic level of photometric scatter.
The data is then binned in two ways. The first
measure is to simply bin the F814W luminosity function in the traditional way.
The second measure is to also bin the data in two dimensions (sometimes referred to as the Hess diagram) the same way as the observations,
using the grid shown in Figure~\ref{WDgrid}. In each case, we then perform a $\chi^2$ fit between model and observed populations to determine the goodness of fit. To completely specify each model, we need to choose values for the
age, the progenitor mass function slope $x$, the distance and reddening, and the two parameters $A$ and $B$ in
equation~(\ref{Mwdfit}).
If we marginalise over these last four parameters, 
we find the best fit (to the Hess diagram) if we adopt the values
$\mu_0=12.05$, $A_{814}=0.37$ and 
\begin{equation}
M_{wd} = 0.502 M_{\odot} e^{0.101 (M - M_{TO})}.
\end{equation}

Keeping these parameters fixed, we show, in 
Figure~\ref{xT2}, the confidence intervals for the fit to age and
main sequence mass function slope $x$ for this improved solution. The solid
contours represent the $1,2$ and $3 \sigma$ ranges obtained by fitting to the
two-dimensional Hess diagram, while the dotted contours result from the fit to
the luminosity function.
The value of $\chi^2_{min}=41.5$ for the Hess diagram fit corresponds to 1.34 per degree of freedom,
as there are 34 total bins used in the fit.
The two sigma age constraint is thus $11.7\pm 0.3$~Gyr for this particular set of parameters. 
The comparison
of the best-fit model ($T=11.61$~Gyr, $x=0.95$) and observations is shown in Figure~\ref{Hessfit2}, and is in excellent
agreement overall. The one obvious disagreement that remains appears to be that the models produce a weaker
turn to the blue than is found in the observations. This can be seen in the F814W$=27.25$ panel
in Figure~\ref{Hessfit2} and also in Figure~\ref{Fit3}, which is a Monte~Carlo realisation of
this best fit model, plotted next to the observations. The location and width of the observed
and modeled sequences are very similar, except for the weaker turn to the blue at the faint
end in the models.

Also shown in Figure~\ref{xT2} are confidence intervals corresponding to a fit of the models to
the observed luminosity function. Although the Hess fit is a more accurate measure, the historical
use of the luminosity function makes this a comparison of interest. Figure~\ref{LF2} shows the
observed luminosity function compared to the best fit model ($x=0.59, T=11.69$~Gyr). Also shown is
the region used to perform the fit. Once again, we sum the bins with F814W$<26$ to use as a single
bin in the constraint. The constraints from the Hess diagram and the luminosity function overlap
considerably, although the luminosity function, not surprisingly, offers a weaker 95\% constraint
on the age ($11.6 \pm 0.6$~Gyr).

\subsection{Internal Chemical Composition}
\label{LowZ}

One advantage the WDCS method has over the MSTO method is that the colours are not metallicity
dependant in any obvious way, which means the subtleties of cluster-to-cluster variations and
of the transformation from observational to theoretical planes is less fraught with peril.
However, the white dwarf method is not completely free of metallicity effects. In particular,
the main sequence lifetime of a given progenitor star is shorter (lower metallicity stars burn
at higher temperatures and hence consume their fuel more rapidly). As a result, the models used
in Hansen et al (2004) likely overestimate the main sequence lifetimes of stars as metal-poor as
those in NGC~6397.

To address this problem, we use the models of Dotter \& Chaboyer (based on the models of
Chaboyer et al 2001), for metallicities 
$[Fe/H] = -2 \pm 0.1$. We have also performed fits of these same models to the main sequence
turnoff (Richer et al 2007), so that our white dwarf cooling age is self-consistent with
the estimates from the MSTO models (at least, to the extent that they are coupled through
the progenitor lifetimes). If we repeat the procedure
of \S~\ref{Default} but using these models instead, the best fit distance, extinction and mass values
are given by  $\mu_0 = 12.05$, $A_{814}=0.36$ and
\begin{equation}
M_{wd} = 0.5 M_{\odot} e^{0.169 (M - M_{TO})}.
\end{equation}
Thus, the best fit distance and extinction are quite robust, although the initial-final mass relation is steeper than before.
The resulting fit for age and $x$ is shown in Figure~\ref{xT4}, yielding a $2\sigma$ range of $11.52\pm0.23$~Gyr from the Hess fit.
The best fit Hess model has $\chi^2 = 39.6$, at T=11.51~Gyr. This is shown in Figure~\ref{Hessfit4}, and a Monte Carlo realisation of this model is shown in Figure~\ref{Fit3_4}.
The best-fit luminosity function (at T=11.46~Gyr) is shown in Figure~\ref{LF4}.
The use of these new models
results in a shift to a slightly lower age (although by less than the difference between
the main sequence lifetimes at fixed mass in the two models) but otherwise the fits are of
similar quality to before, although slightly better in this case.

Thus, the age is reduced somewhat if we use the metal-poor models of Dotter \& Chaboyer, as were used in the
study of the MSTO in this cluster (Richer et al 2007).  
However, the variation between different main sequence
models is also a source of systematic error, so let us now expand our parameter study to use both these models and
those of Hurley, Pols \& Tout (2000), from which our default model was drawn,
 but now with the appropriate metallicity. In this case, we will marginalise over all the
other parameters
including distance, extinction, the initial-final mass relation and also $x$, to obtain a final, all-encompassing constraint on cluster age. The $\chi^2$ curve is shown in Figure~\ref{Tc}, along with the 2$\sigma$ age constraint
$T=11.43 \pm 0.46$~Gyr.

Progenitor metallicities also can result in changes in the cooling models, as the nuclear burning history
will affect the ratio of Carbon to Oxygen (and thus the heat capacity) in the resulting white dwarfs. The
above calculations use the white dwarf compositional profiles from Hernanz et al (1994). We have also performed
 fits using profiles resulting from the models of Hurley et al (2000), but these result in lower $\chi^2$ and
worse fits to the data. Another possibility is that some fraction of the white dwarfs may possess Helium cores (Hansen 2005), as suggested by the strange white dwarf luminosity function of NGC~6791 (Bedin et al. 2005).
 However,a significant population of such
white dwarfs would lead to a peak in the luminosity function at much brighter magnitudes (F814W$\sim 25.7$), which
is not seen in our data. We thus conclude that NGC~6397 contains a negligible population of Helium-core white dwarfs.

\subsection{Hydrogen Layer Mass}

The amount of hydrogen on the surface  of a white dwarf is one of the parameters that has
to be specified for a white dwarf cooling model. The other models in this paper assume a
Hydrogen surface layer with a total mass fraction $q = 10^{-4}$ for each white dwarf. This
is the canonical value expected from standard stellar evolution, but there have been claims
in the past for lower surface values. To that end, we have also performed a fit using
the Dotter \& Chaboyer main sequence models but now with cooling models that have
thinner Hydrogen surface masses, $q_H = 10^{-6}$. The resulting fits yield worse $\chi^2$,
even when we allow the distance, extinction and initial-final mass relation parameters to
float. The minimum $\chi^2 = 55$ ( for an age of 10.55~Gyr), so that we conclude that the
data supports models with a standard hydrogen layer mass.

\subsection{Atmospheric Composition}

In the field, some fraction of white dwarfs show evidence for atmospheres whose dominant
constituent is helium, rather than hydrogen. This has a marked effect on the colours at
faint magnitudes, since helium atmospheres continue to redden as they get cooler, while
hydrogen atmospheres get bluer (in the infrared and optical bandpasses) because of molecular
hydrogen absorption. The good fit of our empirical sequence to the hydrogen model (Figure~\ref{Fit}) suggests
that the contribution of helium atmospheres is small in our case, but we can make this quantitative
by including an extra parameter in our model, the fraction $f_{He}$ of helium atmosphere white dwarfs. However, despite the addition of an extra parameter, this does not result in any improvement
in the fit - essentially because helium atmosphere white dwarfs cool much faster than their hydrogen
atmosphere counterparts and so smooth out the relatively
sharp truncation in the luminosity function or Hess diagram. Furthermore, the colours of cool Helium
atmospheres are redder than for Hydrogen atmospheres, so they do not help to alleviate the slight
remaining discrepancy in the blue hook between model and observations.

\subsection{Binarity}

Binarity can also have an effect on the colours and magnitudes of white dwarfs. A white
dwarf in a binary with a main sequence star will be overwhelmed and excluded from our sample,
but two white dwarfs in a binary will add together and potentially be brighter than a single
isolated star. We model this binarity by randomly selecting the progenitors from the initial
mass function and then accounting for the evolution of the two white dwarfs as isolated objects,
eventually summing their light in the model. The resulting simulated samples result in degraded
$\chi^2$, indicating that there is little to no evidence for binarity in our white dwarf sample.

To illustrate this, we show in Figure~\ref{Bin} the effect of a given binary fraction on the
$\chi^2$ per degree of freedom for a model with all the other parameters held fixed at the
best fit values for the Z=0.001 model. The open circles represent fits to the luminosity
function and the filled triangles represent fits to the Hess diagram. We note that the luminosity
function constraints are much more strict. This is because binarity results in model stars being found
in some bins in the grid where they are never found otherwise (and therefore we don't usually use
these bins in our $\chi^2$ fits because they contain no information). Thus, the open circles 
represent Hess fits with an expanded diagram (which feature three more bins at the bright, red
end of the grid in Figure~\ref{WDgrid}). This results in much stronger constraints, and we use
this to quote a 2$\sigma$ upper limit on the binary fraction $<4\%$. This represents the fraction
of white dwarfs that have another white dwarf as a companion.

\subsection{The Main Sequence Turnoff}

The traditional method for determining globular cluster ages is to fit stellar models
to the main sequence turnoff. In Richer et al (2007) we compare our observations of the
turnoff in this cluster to four sets of theoretical isochrones. In each case, good fits are
obtained for a wide variety of ages, $12 \pm 2$~Gyr, where the accuracy is limited, as in
prior studies, 
by the uncertainties in the distance and extinction. When the distance and extinction used
in the MSTO method are restricted to the range that fits the white dwarf model (see \S~\ref{Distance})
and ages determined using the models of Dotter \& Chaboyer, we find a best fit age of 11.6~Gyr,
with a 95\% confidence lower limit of 10.6~Gyr. Hence the two methods are in complete agreement,
as these were the models used to determine the progenitor ages in Figure~\ref{xT4}.
As a result, this combination represents a successful end-to-end (main sequence to white dwarf) comparison of a set
of stellar models with the observed stellar population.

\section{Discussion}
\label{WDDiscuss}

The fits obtained here from the white dwarf cooling sequence result in a constraint
that is quite a bit tighter than the traditional MSTO method. The reason for this
comes from the fact that we have modeled the entire cooling sequence, rather than
just a localised feature (as in the case of the turnoff). In principle,
simply fitting a cooling model to the truncation in the observed cooling sequence leads
to a similarly simple constraint, but it is subject to several unsatisfactory
parameter degeneracies, including those between age, white dwarf mass and 
cluster distance. By constructing a self-consistent model for the
entire cooling sequence, these degeneracies are lifted because different choices of
white dwarf mass and age have different consequences for distributions of white dwarfs
in both colour and magnitude. We illustrate this by performing such a simple fit in
appendix~\ref{Simple} and contrasting it with our more detailed procedure.

To arrive at a final age constraint, we have marginalised over uncertainties in distance,
extinction, the white dwarf mass (and it's variation along the WDCS) and binary fraction, as well as model
systematics that result from different choices of main sequence models, white dwarf
internal compositions and hydrogen layer
masses. We have restricted our models to those of appropriate metallicity (so the models in
\S~\ref{Default} are not included in the final fit) and thus our final age constraint is
that derived from Figure~\ref{Tc},
$T=11.47 \pm 0.47$~Gyr.

To further understand which information is important in determining this constraint, let us
consider the nature of the best fit models.

\subsection{The Best-Fit Solution}

The $\chi^2$ value per degree of freedom for the best fit models is $\sim 1.27$, so it
is good, but not perfect. Figure~\ref{Hessfit4} shows the level of agreement between the
data and the best fit model. The model reproduces well not only the colour distributions at each
magnitude, but also the normalisations at different magnitudes relative to each other.
The only feature of the observations that is not well reproduced by our models is the
blue colours at magnitudes F814W$\sim 27.25$. This can also be seen in the visual comparison
of the two panels in Figure~\ref{Fit3_4}, where the models clearly do not fully reproduce the clump
of stars bluewards of the faint end of the main locus. This most likely indicates some level
of mismatch between the theoretical colours and the true colours at the faintest temperatures.
The deviations are in the same sense as those seen in Figure~\ref{Fit}, where the model colours
lie a little lower than the empirical cooling sequence at the faint end. The fact that the blueward shift
of the colours is driven by collisionally induced absorption of molecular 
hydrogen (Mould \& Liebert 1978; Bergeron, Saumon \& Wesemael 1995; Borysow, Jorgensen \& Zheng 1997; Hansen 1998; Saumon \& Jacobsen 1999)
suggests that these deviations may 
 indicate residual deficiencies in
the models or simply that the atmospheric composition contains an admixture of Helium along with the Hydrogen.
 Nevertheless, we emphasize again that
the model fits are still good despite this residual mismatch and that improving the colours is likely to only make the
age constraint even tighter. 

Let us now consider the properties of the underlying white dwarf model.
Figure~\ref{MM} shows the mass of the white dwarf as a function of F814W magnitude
for the best-fit solution. Note, the F814W magnitude shown in this plot does not include the
photometric scatter. In the upper panel we show the observed luminosity function,
to indicate the range over which the cutoff occurs. The white dwarfs in this part
of the luminosity function range from 0.52$M_{\odot}$ to $0.62 M_{\odot}$. This is
important, because it means the more massive white dwarfs do not contribute significantly
to the luminosity function. This is because they cool faster at late times (as can be
seen by the flattening in the F814W-M$_{\rm wd}$ curve). A consequence of this is that we
cannot directly interpret the cluster age as the white dwarf cooling time, because even
the white dwarfs at the start of the truncation come from relatively low-mass main sequence stars, and
therefore have a small contribution to the total age from the main sequence
lifetime. This is shown explicitly in Figure~\ref{MM2} where we show a similar diagram
to Figure~\ref{MM} but we now show the main sequence mass, related to the white dwarf
mass through the initial-final mass relation. Also shown is the corresponding main sequence
lifetime from the Dotter \& Chaboyer models. We see that the truncation in the white dwarf
luminosity function appears when the main sequence age of the progenitors starts to drop
precipitously, so that we get older, cooler and fainter white dwarfs. Over a range of only
0.5 magnitudes, we see that the cooling time of the white dwarfs at these magnitudes changes
by $>5$~Gyrs! This is the ultimate source of the truncation -- a bin of fixed width in magnitude
corresponds to a much smaller range in the progenitor mass function than higher up the
cooling sequence, and so there are fewer stars in that bin. The underlying physical reason for
this is that more massive white dwarfs initially cool more slowly because they have larger
heat capacity, but they crystallise earlier because of the high central densities and thereafter
cool more quickly as the heat capacity enters the Einstein-Debye regime. This means that, at any
given age, there is a `leading edge mass', above which the white dwarfs have overtaken their
less massive counterparts, and cooled beyond detection.

We can now also return to the issue of how the luminosity function truncation is
affected by incompleteness. Figure~\ref{IncOff} shows the luminosity function
taking into account the observational incompleteness (solid histogram) and the
corresponding complete luminosity function (dashed histogram). We see that the
sharp drop is a direct consequence of the model, and not strongly affected by
observational incompleteness, even at F814W$ \sim 28$, where the recovery
fraction is only $53\%$. The reason for this is that we lose white dwarfs primarily 
through confusion with brighter stars, not because of large photometric scatter from noise 
in those that we do recover. To understand this better, consider Figure~\ref{Spread},
in which we show the results of artificial star tests for stars inserted with F814W=27.9.
The fraction of stars never recovered at all is 47\%, but the dispersion of those that
are recovered is only $\sim 0.1$ magnitudes. This is not sufficient to alter the shape
of the recovered luminosity function.

As a final demonstration of the properties of our modelling procedure, we show in
Figure~\ref{Fit4} the effect on the model population as we change the age. The middle
panel shows the best-fit model from Figure~\ref{Fit3_4} and the left and right panels show
the same model but for ages of 10~Gyr and 13~Gyr, with all other parameters held fixed.
The younger population is clearly distinguishable from the best-fit model, with a truncation
at brighter magnitudes (the $\chi^2=171$ for this model). The population at 13~Gyr shows no
abrupt edge except that due to incompleteness at F814W$>28$,
 and a much stronger colour evolution ($\chi^2=185$). 

\subsection{Distance Scale and Extinction Redux}
\label{Distance}

The shaded region in Figure~\ref{MuA} shows the 2$\sigma$ range for NGC~6397 distance modulus and
reddening, for the fit to the Dotter \& Chaboyer models. This is compared to the equivalent
values from three studies determined by comparing the main sequence with field M-subdwarfs measured
by Hipparcos.
The solid point is the value obtained by Reid \& Gizis (1998), which is probably the most
easily compared to our value, since the comparison was made in the V \& I bands. Indeed, our
value is consistent with the low end from that study. The value obtained by Reid (1998), used
the B \& V bands, is shown as the open circle, and is somewhat higher than our white dwarf distance.
The best agreement is obtained with the value quoted by Gratton et al (2003), obtained again using
the B \& V bandpasses, but with a variety of assumptions regarding colour corrections etc that differ
from those of Reid. 
In fact, our distance is somewhat better constrained than the main sequence values,
 yielding $\mu_0 = 12.03 \pm 0.06$ at
$2 \sigma$, while the reddening is constrained to be $E(F606-F814)=0.20\pm 0.03$.

\subsection{Initial-to-Final Mass Relation}
\label{IFMRsec}

One of the inputs into our model is the relationship between the white dwarf mass
and the mass of its progenitor. This is a fundamental quantity that has been studied
for many years, so it is natural to ask how our best model fits correspond to those
determined empirically. Figure~\ref{IFMR} shows two of our fits (using our default
models and then the metal-poor Dotter \& Chaboyer fits in \S~\ref{LowZ}) compared
to a well-known empirical IFMR from Weidemann (2000) and a more recent fit from
Ferrario et al (2005). It is encouraging that our best fit model recovers a relation
similar to the empirical ones, although it must be noted that our fit really only
probes a limited range of masses, as inferred from Figure~\ref{MM}, essentially 
the unshaded region in Figure~\ref{IFMR}. We note also that our IFMR is in agreement
with the results of Moehler et al (2004), who found, using multicolour photometry, that
the mean mass of white dwarfs at the top of the cooling sequences in NGC~6397 and NGC~6752
was $0.53 \pm 0.03 M_{\odot}$. This is also in agreement with theoretical expectations
(Renzini \& Fusi Pecci 1988).\footnote{Renzini \& Fusi Pecci make a second prediction which
could, in principle, be constrained by our results. They predict a very small dispersion in
the white dwarf mass $\sigma_M \sim 0.003 M_{\odot}$, which should translate into a dispersion in
magnitude at fixed colour on the upper part of the cooling sequence. Our measured dispersion
for white dwarfs with F814W$<25$ is $\sigma_{814}=0.095$ (measured as the dispersion about the
best fit linear trend on the upper cooling sequence). This translates to a variation in white
dwarf radius of $\delta R/R = 0.044$, and so $\sigma_M \sim 0.066 M_{\odot}$. It is an order of magnitude
larger than the prediction and larger than the scatter in the photometry at these magnitudes. 
We might infer a larger dispersion in mass than expected, although there are other possible contributions
such as 
variations in hydrogen layer mass and spectral composition (DB stars have smaller radii). It is worth
noting though, that this width is narrower than the estimated width of the field sample $\sigma_M = 0.137 M_{\odot}$ (Bergeron, Saffer \& Liebert 1992).}



\subsection{Cosmological Considerations}

One of the more exciting aspects of this result is that we get a finite upper
limit on the age. Most other measurements, including our M4 result, yield primarily lower
limits on the age of star clusters when the full systematics are taken into
account. The reason for the excitement is that we can now start to properly
constrain the cosmological epoch at which NGC~6397 formed.

Let us consider a flat universe with a cosmological constant and parameterised
by the present day matter density $\Omega_0$. The present-day age for an object
formed at redshift $z_c$ is (Weinberg 1989)
\begin{equation}
T (z_c) = \frac{2}{3} \left[ 1 + \frac{\Omega_0}{1 - \Omega_0} \right]^{1/2}
H_0^{-1} \left( \sinh^{-1} \left[ \frac{1 - \Omega_0}{\Omega_0} \right]^{1/2}
- \sinh^{-1} \left[ \left(\frac{1 - \Omega_0}{\Omega_0}\right)^{1/2} (1 + z_c)^{-3/2} \right] \right)
\label{TZ}
\end{equation}
where $H_0$ is the present-day Hubble constant (which we will take to be 71~$\rm km.s^{-1}.Mpc^{-1}$).
Using the parameters from the best fit cosmologically flat model to the microwave background anisotropy (Spergel et al 2003),
we take 
 $\Omega_0=0.27$, which gives a present day age of 13.7~Gyr for the universe. We
can use equation~(\ref{TZ}) to figure out the range of possible formation redshifts, given the
age of NGC~6397, which we  take to be $11.47 \pm 0.47$~Gyr (our $2 \sigma$ value). The resulting
formation redshift is 
\begin{equation}
z_c = 3.1 \pm 0.6.
\end{equation}
Figure~\ref{Age} shows this result compared to the relation (\ref{TZ}) for the WMAP cosmology.
Our result places the formation epoch of NGC~6397 somewhat more recently than
the reionization epoch (generally taken to be at redshifts of 6 or greater) but quite naturally
associated with the copious star formation seen at redshifts $\sim 2-4$. 
As noted above, we are in agreement with 
 prior age estimates for the Globular Cluster system based on the
MSTO method  but those estimates could only
place a lower limit on the age of the system. The greater precision of our method now
allows us, for the first time, to determine a realistic offset between the creation of
the universe and the formation of the globular cluster system, while also confirming the significant delay
between globular cluster formation and the onset of star formation in the Galactic disk
that we found in Hansen et al. (2002).

The age for NGC~6397 places it's origin firmly near the peak of the cosmic star formation
rate, as measured by various deep cosmological surveys (Madau et al 1996; Thompson et al 2006), shown
in Figure~\ref{SFR}. 
The origins of globular clusters has been a subject of debate for at least forty
years. The first proposed scenarios postulated a very early origin (Peebles \& Dicke 1968; Fall \& Rees 1985) although the possible role of mergers was also recognised (Searle \& Zinn 1978; Schweizer 1987; Ashman \& Zepf 1992). The position of globular clusters within the modern hierarchical
structure paradigm is highly uncertain, and our age determination now offers the opportunity to
draw a distinction between at least some of the competing proposals. With a formation redshift
$z \sim 3$, NGC~6397 appears to have formed after the epoch of reionization, which is conservatively held to be at $z>6$ (Fan, Carilli \& Keating 2006). If we assume NGC~6397 is representative of the class of
metal-poor globular clusters as a whole, this rules out the notions that cluster formation was triggered by
radiation pressure-driven collapse in protogalactic haloes during reionization (Cen 2001) or that
the clusters were themselves responsible for the reionization (Riccotti 2002). Furthermore, 
the age of NGC~6397 seems inconsistent with the notion of
Forbes et al (1997) that the metal-poor clusters formed early and were `switched off' by some
later event, possibly reionization (Santos 2003). Our results are much more consistent with
scenarios that associate cluster formation with the epoch of rapid, starburst-driven star
formation that occurs at $z \sim$ 2--3, be it by gas collapse in dense, gaseous disks of
high redshift galaxies (Kravtsov \& Gnedin 2005) or collapse driven by starburst winds compressing
the gas in accreting satellites (Scannapieco, Weisheit \& Harlow 2004).

Of course, we base these inferences on a single cluster. The role of NGC~6397 as a tracer of
the earliest star formation epoch in our galaxy is based on its low metallicity. The orbit, however, is not exactly what one might expect from a true halo object (Milone et al 2006; Kalirai et al 2007), never getting more than 3~kpc above the galactic plane. Although MSTO methods place the
ages of the outer globulars as consistent with that of NGC~6397 (e.g. Salaris \& Weiss 2002) the
$2\sigma$ accuracy of the method ($\sim 1.5$~Gyr) means the constraint is much less stringent than
our absolute age measurement.

\subsection{Outlook}

The age constraint we derive is more precise than those of previous studies using the MSTO method,
and also more precise than our earlier result on M4 (Hansen et al 2004). In our M4 study we also
tried to calculate ages using other white dwarf models from the literature. We have not done so in
this study because we now require knowledge of the models in more detail than can be gleaned from
published material. We hope that other groups will perform similar analyses using their models, so
that we may gain a more realistic understanding of the systematic uncertainty in the age based on
using different evolutionary codes. 
 We caution, however, that simple heuristic, comparisons between
models and observations 
are unlikely to yield particularly interesting answers, as can be seen from appendix~\ref{Simple}. To aid in proper comparison, we include
in Tables ~\ref{DI} and \ref{DV} the results of our artificial star tests, in which we quantify the amount
of photometric scatter that results as a function of intrinsic magnitude. Feeding theoretical models through this 
matrix and then comparing to the data in Table~\ref{CMDtable}, one can perform statistically well-posed fits between
models and data.

The prospects of extending this methodology to other clusters is somewhat limited by the extreme faintness
of the coolest white dwarfs. The only other clusters for which this seems feasible are NGC~6752 and 47~Tucanae.
Nevertheless, similar observations of these clusters could be very illuminating, as this trio represent
the archetypes for the metal poor, intermediate metallicity and metal-rich/thick~disk/Bulge cluster families.
In particular, NGC~6397 and NGC~6752 are considered contemporaneous by MSTO age determinations, while
47~Tuc may be marginally younger (e.g. Sarajedini, Chaboyer \& Demarque 1997; Rosenberg et al 1999; Gratton et al 2003). The white dwarf cooling sequence offers us then the chance to test these assertions at greater accuracy. However, the observational expense involved in such projects will prevent similar tests in more distant
clusters.

One other important outlook is to confirm our method of background subtraction by repeating
this analysis with a true proper motion-selected sample, all the way down to the cluster cooling
sequence truncation. This will be performed in the coming year, in HST program GO--10850. While we are
 confident that the gross features of the white dwarf population as outlined here will remain
the same, a proper motion selection will shed important new light on the less populated parts of
the colour-magnitude diagram. In particular, we anticipate the discovery of a tail of higher-mass white dwarfs fainter
and bluer than the bulk, which are presently confused with the residual galaxy population. Such a tail
will help us to understand whether the residual differences between the observed and model populations are
a consequence of a colour mismatch or a missing piece of physics.

\section{Conclusion}

In conclusion, we have performed a detailed Monte-Carlo simulation of the white dwarf
population in the metal-poor globular cluster NGC~6397, comparing it to the observed
WDCS obtained with the ACS camera on HST. The principal conclusion that results from this
comparison is that the age of the cluster is  $11.47 \pm 0.47$~Gyr at 95\% confidence.
NGC~6397 is a member of the class of metal-poor clusters that are thought to be amongst the
oldest objects in the Milky Way and so this age places the epoch of original assembly of the 
Galaxy at $z =3.1 \pm 0.6$. 

Our model is also in agreement with various independent measures of several of the parameters,
such as the distance and extinction along the line of sight, and the relationship between
white dwarf mass and progenitor mass.

\acknowledgements
BH, JA, RMS, IK \& MMS acknowledge support from proposal GO-10424, and
JK is supported through Hubble Fellowship grant HF-01185.01-A,
 all of which were provided
by NASA through grants from the Space Telescope Science Institute, which
is operated by the
 Association of Universities
for Research in Astronomy, Inc., under NASA contract NAS5-26555.
HBR also acknowledges the support of
the US-Canada Fulbright Fellowship Committee.
The research of HBR \& GGF is supported in part by 
the  Natural Sciences and Engineering Research Council of Canada.


\appendix

\section{A Simple fit to the Cooling Sequence Truncation}
\label{Simple}

In this appendix, we perform a simple-minded comparison between our cooling models
and the observed truncation of the white dwarf sequence. Although our full model
comparison results in much tighter constraints, this exercise is still instructive 
as it further illustrates how some of the ancillary information is used in the
proper models.

As a first step, we characterise the magnitude of the truncation as F814W$=27.6\pm0.1$.
Using the distance and extinction from Reid \& Gizis (1998), we convert this into an
absolute magnitude ${\rm M_{814}}=15.15 \pm 0.15$. We can thus compare this to a model
to infer an age. However, we first need to specify the mass of the white dwarf, and this
is a free parameter in the absence of other constraints. Figure~\ref{Curve} shows the
absolute magnitude as a function of age of Hydrogen atmosphere models of various masses. A further constraint
is included, requiring that the F606W-F814W colour of the models lie in the range $1.13\pm 0.13$
(the dereddened colour range spanned by the observations at the location of the truncation).
 Those parts of the curve that meet
this criterion are solid, while those that do not are dashed. We see that acceptable ages
are found for all masses, the cooling age increasing for masses from $0.5 M_{\odot}$ to $0.65 M_{\odot}$
and then decreasing thereafter (because of the faster cooling -- due to core crystallisation -- of
more massive white dwarfs at late times). Thus, in the absence of other information, we infer
an age range $10.2 \pm 1.4$~Gyr (cooling age only). 

The reason that the full models yield a precise value is because they incorporate other constraints.
Most notably, the models have to fit the location of the cooling sequence at the bright end (which favour
models in which the white dwarfs have 
masses $<0.55 M_{\odot}$ initially). With this additional constraint, the shape of the observed cooling
sequence constrains how much the mass can vary along the cooling sequence and the manner in which the
numbers increase along the sequence constrains which cooling models and which masses yield good fits.
Finally, this full modelling approach also incorporates the main sequence lifetime, which contributes
about 5-10\% to the final age for stars at F814W=$27.6$ in the best fit model (this can be seen by
comparing Figures~\ref{MM2} and \ref{Curve}.)

Thus, we can build a rough picture of where the final number comes from. Our best-fit models suggest
the white dwarf mass at the truncation is $\sim 0.6 M_{\odot}$, which corresponds to a cooling age
of $\sim 11.0 \pm 0.5$~Gyr, according to Figure~\ref{Curve}. The initial-final mass relation emerging
from the fit yields a progenitor mass $\sim 2 M_{\odot}$, which in turn yields a main sequence lifetime
$\sim 0.5$~Gyr. Thus, we arrive at a cluster age $\sim 11.5 \pm 0.5$~Gyr.

\section{Proper Motions}

Approximately 54\% of our ACS field has some prior WFPC2 imaging, so proper motion
separation is, in principle, possible. However, the first epoch exposures are not
deep enough to allow us to perform proper motion separation at a level necessary to
reach the end of the white dwarf cooling sequence.

Figure~\ref{IRO} shows the proper motion displacement ($\mu$, in units of ACS pixels) for all 
point sources (all objects in Figure~\ref{pretty_cmd}) as a function of F814W magnitude. 
This is done by matching each ACS detection to the closest 2~$\sigma$ peak in the 
WFPC2 data. This works well for magnitudes F814W$<26$, leading to a very clear separation
of the cluster ($\Delta \mu < 1.5$ pixel) and Galactic field stars ($1.5<\mu<5$ pixels). However,
we can see that, at magnitudes F814W$>26.5$ there is a very incomplete separation of the
cluster stars. There is clearly still a clump at small $\mu$, but the distribution of
$\mu$ is large and consistent with noise.

\section{Cluster Dynamical Effects}
\label{Jarrod}

NGC~6397 is a post core-collapse cluster. At the high central densities experienced in
this cluster, many exotic dynamical interactions can occur between stars. Stars in binaries
can be ejected, single stars can be exchanged into binaries and mass transfer between stars
can be either halted or initiated. It is thus natural to wonder whether such processes have any
effect on our parameter estimation. 

The first thing to note, of course, is that our field is chosen to minimise contributions 
of this kind, since it is located well away from the high density core. Nevertheless, mass
segregation is an important process in globular clusters and some stars with exotic
evolutionary histories may be able to migrate
outwards far enough to enter our sample. To examine this possibility, we have examined an N-body 
simulation of NGC~6397, calculated in the manner described in Shara \& Hurley (2006). The simulation
began with 100,000 stars, 5\% of which were in binaries initially. The model cluster reached core
collapse at 15~Gyr, with 25,000 stars remaining, so its late-time structure is qualitatively similar
to that observed for NGC~6397.
 Our modeling procedure is already able to treat a binary
population and our data also show little or no evidence for binarity. The bigger worry is
that we are not modeling the age distribution of white dwarfs properly because of the
contribution of so-called 'divorced' white dwarfs -- stars that spent some time in a binary,
experienced mass transfer which altered the stellar evolution 'clock', and were then
removed from the binary by a dynamical interaction. It is these stars that are not necessarily
so easy to identify directly, since the only way they will stand out from the regular cooling
sequence is if there is a sufficiently large mass difference with similar
age white dwarfs from single star evolution (`bachelor' white dwarfs) to make a measureable 
difference in the photometry. Given the telescoped nature of the initial-final mass relation,
this is not a very sensitive test.

The upper panel of Figure~\ref{2Pan} shows the radial distribution of bachelor and divorced white dwarfs from
the Hurley \& Shara simulation at an age of 13~Gyr. We see that the divorced white dwarf
profile is essentially a scaled version of the bachelor radial profile. The scaling factor 
is $\sim 7\%$ i.e. at any radius, the divorced white dwarfs make up roughly 7\% of the white
dwarfs. The lower panel shows the age distribution of the divorced white dwarfs (points)
compared to a scaled version of the bachelor age distribution. Once again, there is excellent
agreement, suggesting that the degree of dynamical interaction and mass transfer is not
sufficient to dramatically alter the stellar evolution clock. The one potential discrepancy
is a 2$\sigma$ excess of white dwarfs with $T_{wd}<200$~Myr i.e. divorced white dwarfs
that were born anomalously early. However, even if this is real, the excess corresponds to
roughly 5\% of the divorced white dwarfs, which are themselves a minority population. Overall,
this corresponds to roughly 3 anomalously young white dwarfs per sample of 1000. This is far
too little to have any effect on our results.

\clearpage
\plotone{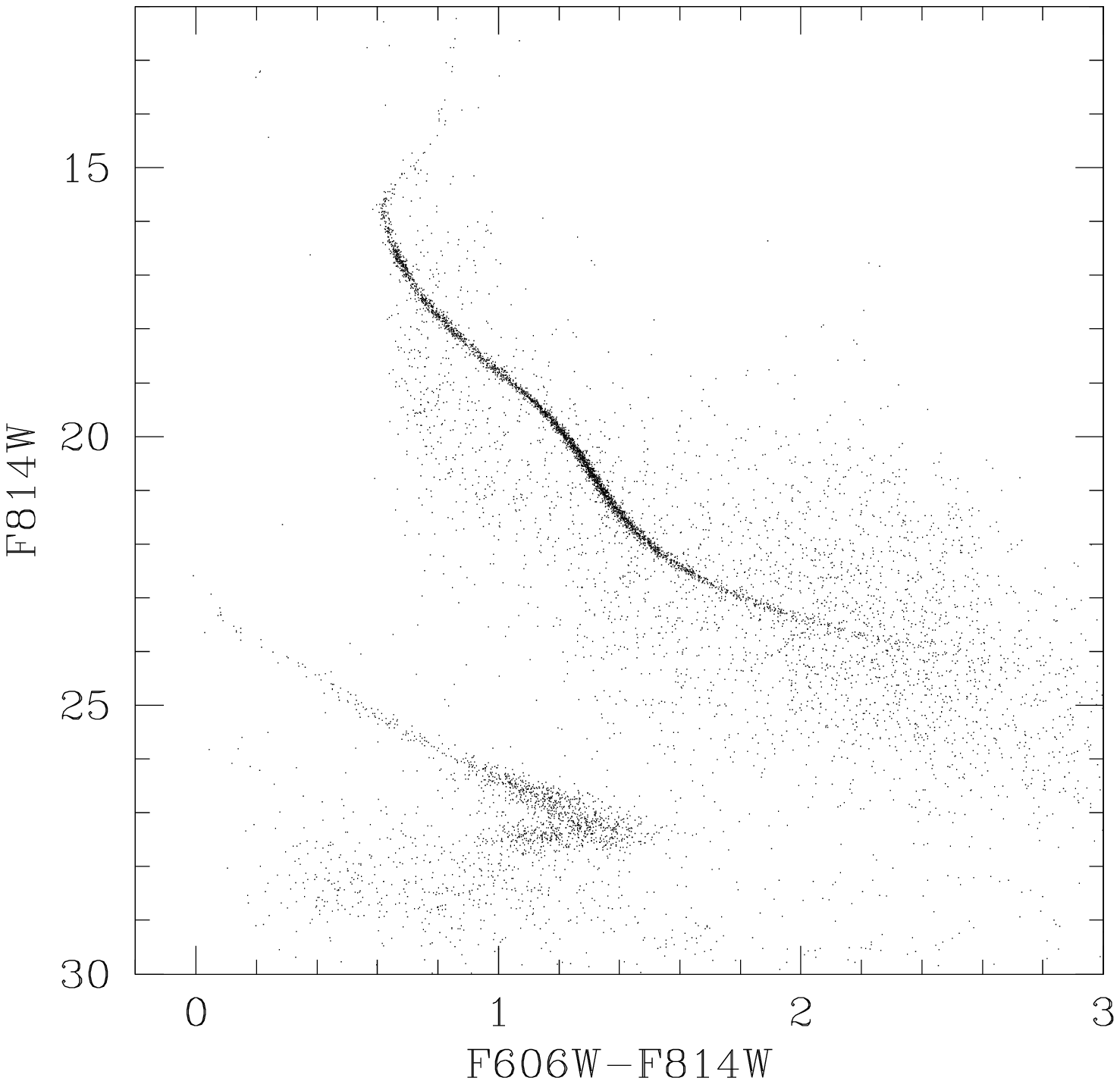}
\figcaption[CMD2.ps]{ The ACS colour--magnitude diagram for our field in NGC~6397.
All real point sources are shown (so the extended galaxy population is not shown).
 Prominent features include a
cluster main sequence, a clear Main sequence turnoff and a clear white dwarf
cooling sequence. Most important is the clear evidence for a sharp decline in
the number of white dwarfs at magnitudes greater than F814W$ = 27.6$. The detectability
of sources at fainter magnitudes is evident from the fainter, bluer population of
background galaxies that survive the point source cuts.
 \label{pretty_cmd}}

\clearpage
\plotone{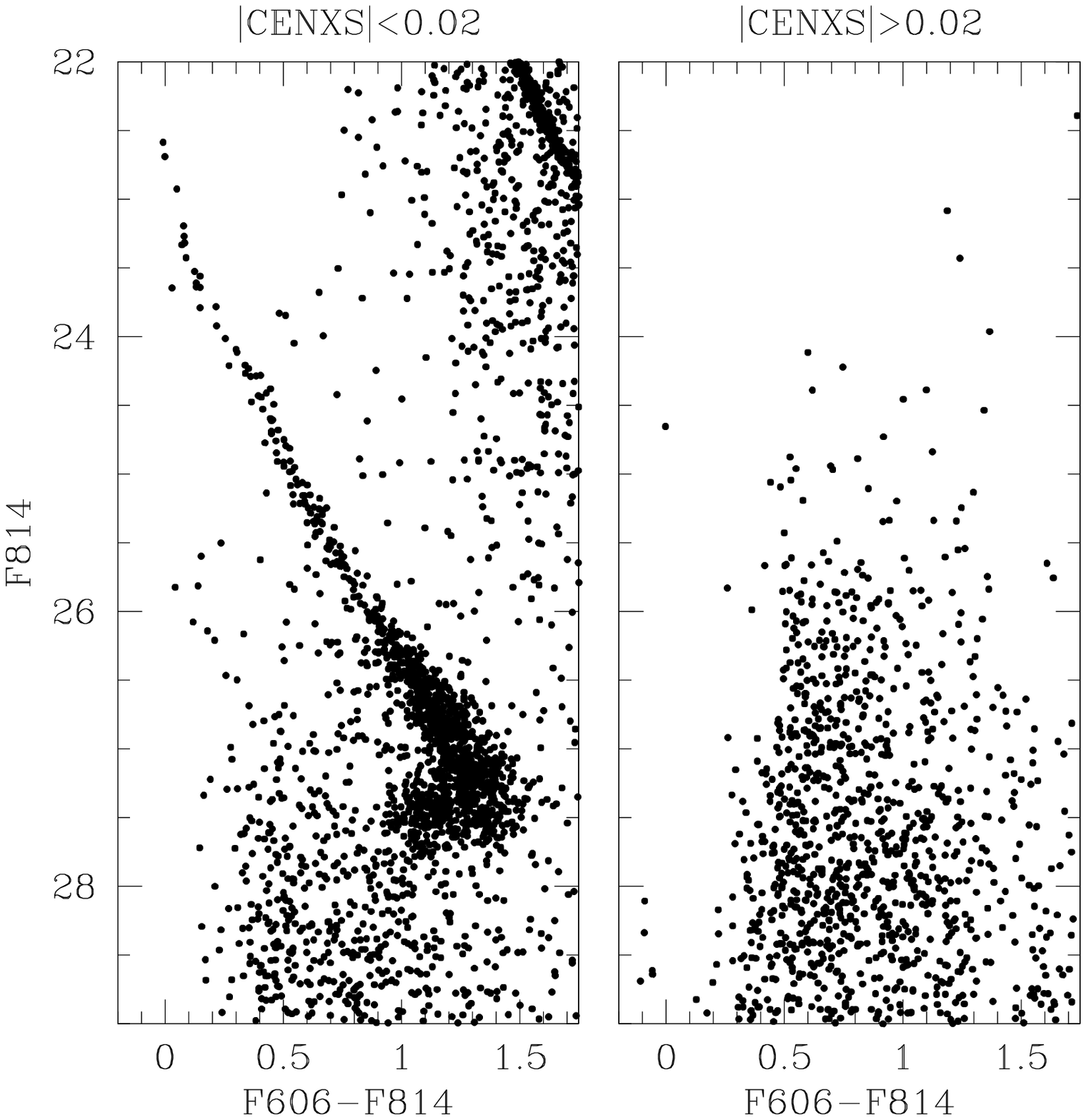}
\figcaption[WDs.ps]{The left hand panel shows the point sources in the region of the
colour magnitude diagram that encloses the white dwarf population. The right hand
panel shows the full population of extended sources in the same region -- this is the
background galaxy population.
\label{pretty_wd}}

\clearpage
\plotone{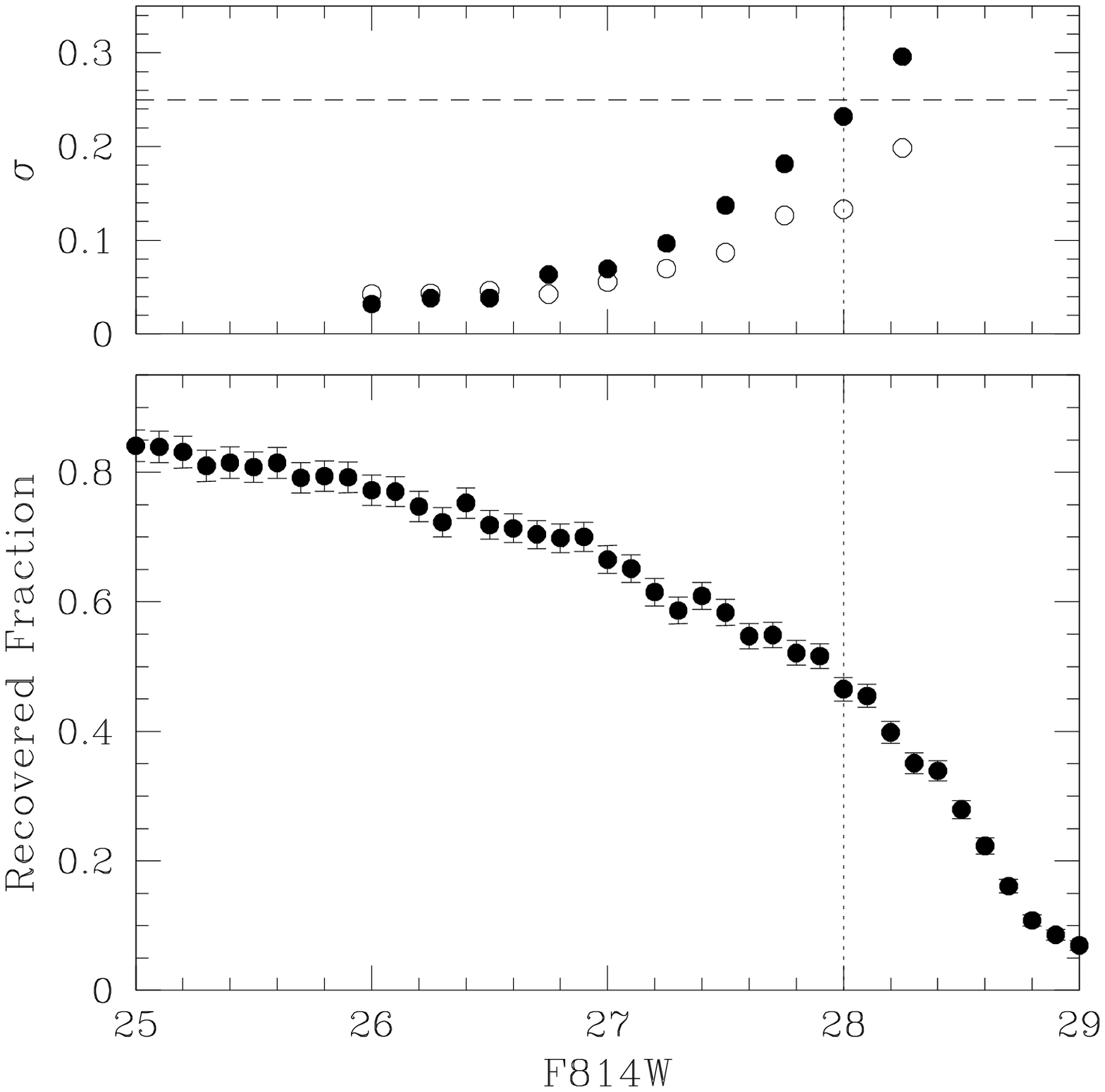}
\figcaption[f3.ps]{ The lower panel shows the recovery fraction of artificial white dwarfs as a function
of F814W magnitude. The upper panel shows the dispersion in the recovered magnitudes. The distribution is asymmetric with respect to stars recovered at fainter/greater magnitudes (solid points)
and brighter/smaller magnitudes (open points) and so the dispersions are fit independently.
The vertical dotted line indicates an input magnitude F814W=28 and the dashed line shows a
magnitude of 0.25 mags.
\label{Rec}}

\clearpage
\plotone{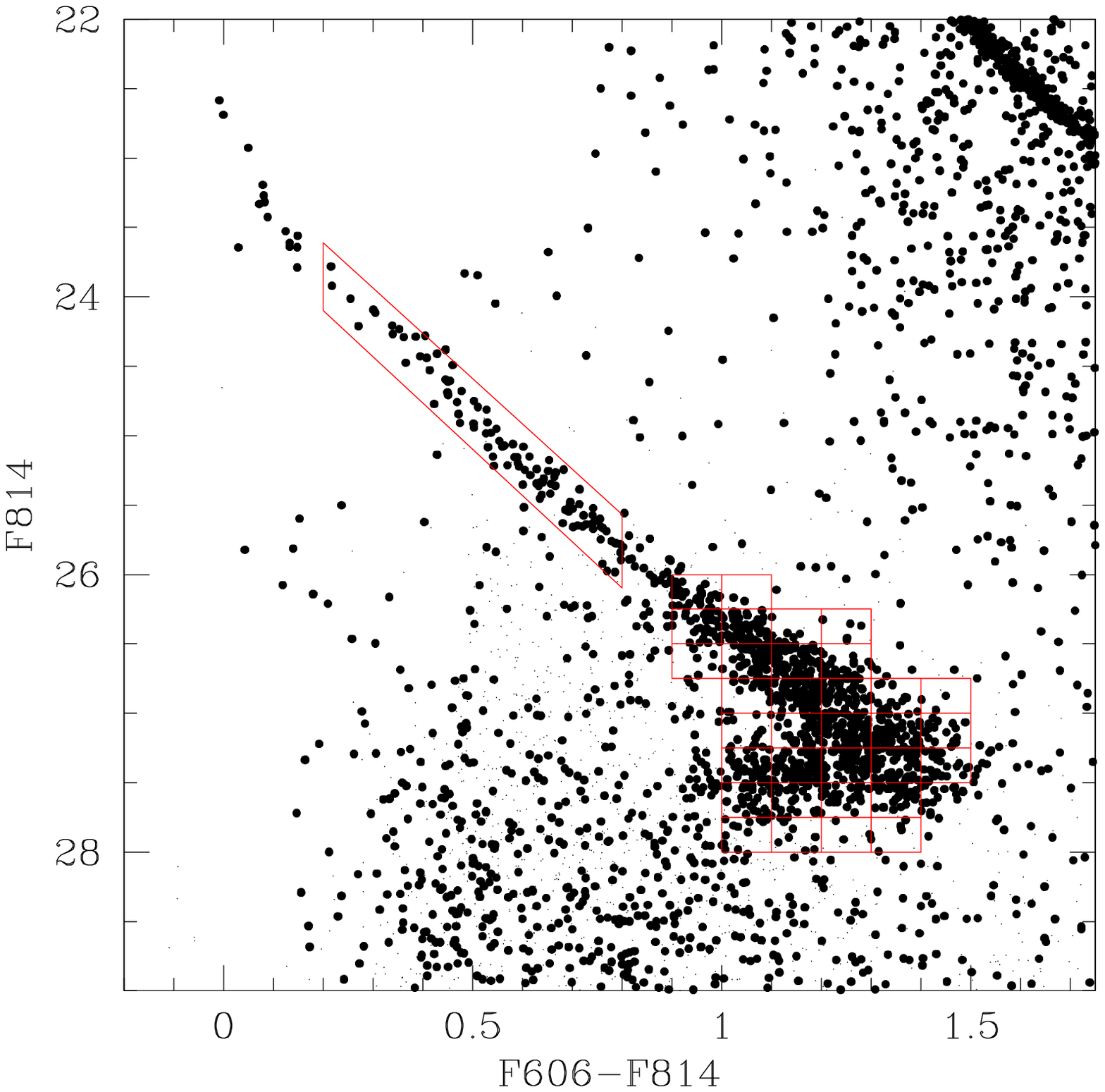}
\figcaption[WDgrid.ps]{The red grid overlaid on the white dwarf population indicates
the manner in which we will bin the results for comparison to the models. The
small points show the location of extended sources.
\label{WDgrid}}

\clearpage
\plotone{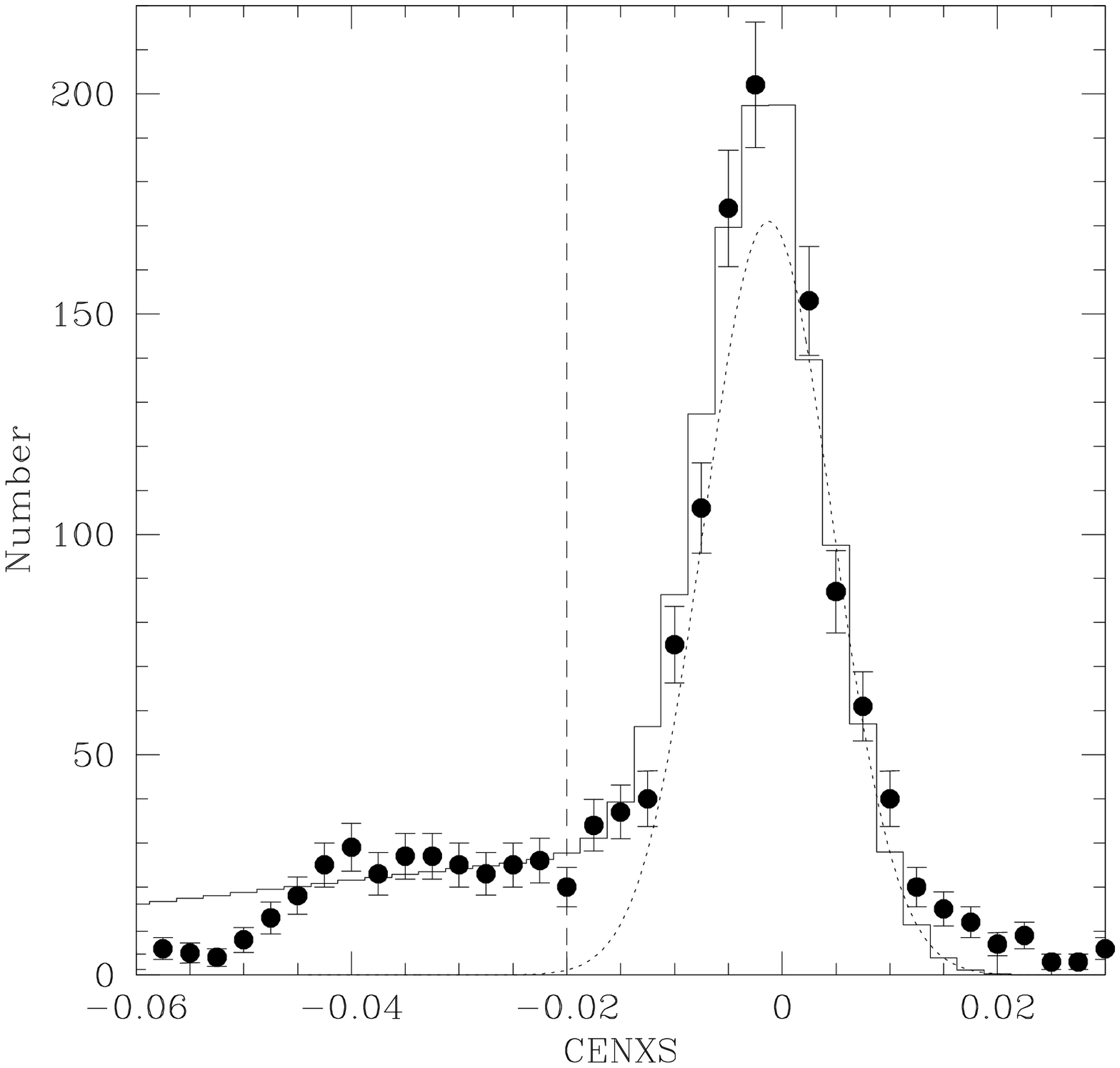}
\figcaption[Mod.ps]{The solid points indicate the distribution of the {\tt CENXS} parameter for all
objects that fall within the grid we use to analysis the white dwarf luminosity function.
The dotted curve indicates the best fit gaussian distribution to the main peak of stellar
objects. The solid histogram includes both this and a second (one-sided) gaussian  to represent the
partially resolved galaxy contribution.
\label{SharpDis}}


\clearpage
\plotone{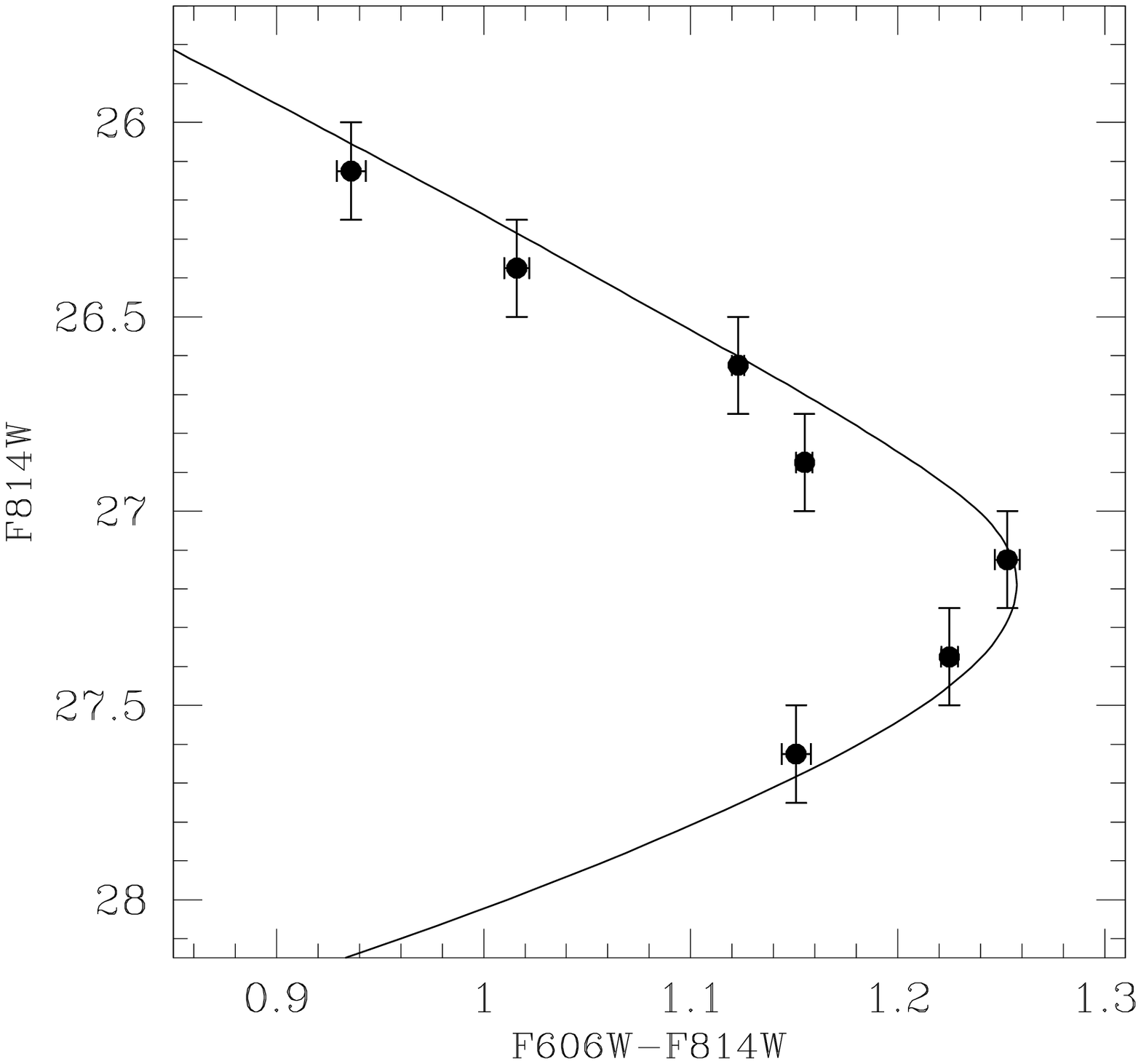}
\figcaption[Fit0.ps]{The large solid points are our empirical cooling sequence, as derived in the
test. The solid
line is the F606W-F814W/F814W cooling sequence for fixed radius from the models of Bergeron,
suitably reddened and shifted, for a $0.5 M_{\odot}$ model. The reddening used here is E(F606W-F814W)=0.16
and the vertical shift is $\mu_{F814W}=12.3$.
\label{Fit}}

\clearpage
\plotone{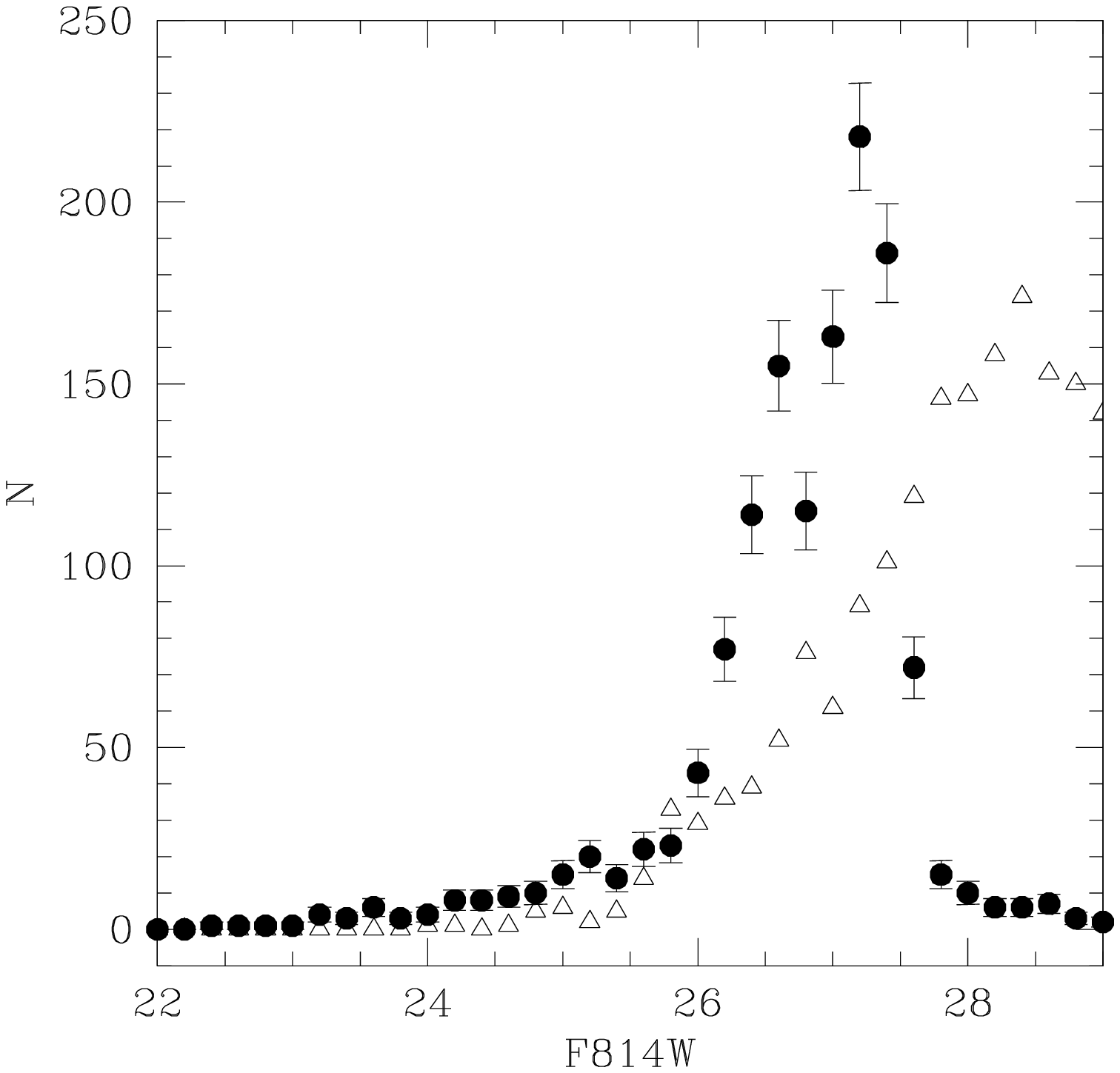}
\figcaption[LF0.ps]{The solid points indicate the luminosity function of white dwarfs,
identified by cuts of $\left| \right. ${\tt CENXS}$\left. \right|<0.02$ and {\tt ELONG}$<0.02$.
The open triangles are the luminosity function of all objects bluer than the WDCS. The smoothly
rising luminosity of these (mostly) galaxies argues strongly that the observed truncation in
the luminosity function is real.
\label{LF0}}

\clearpage
\plotone{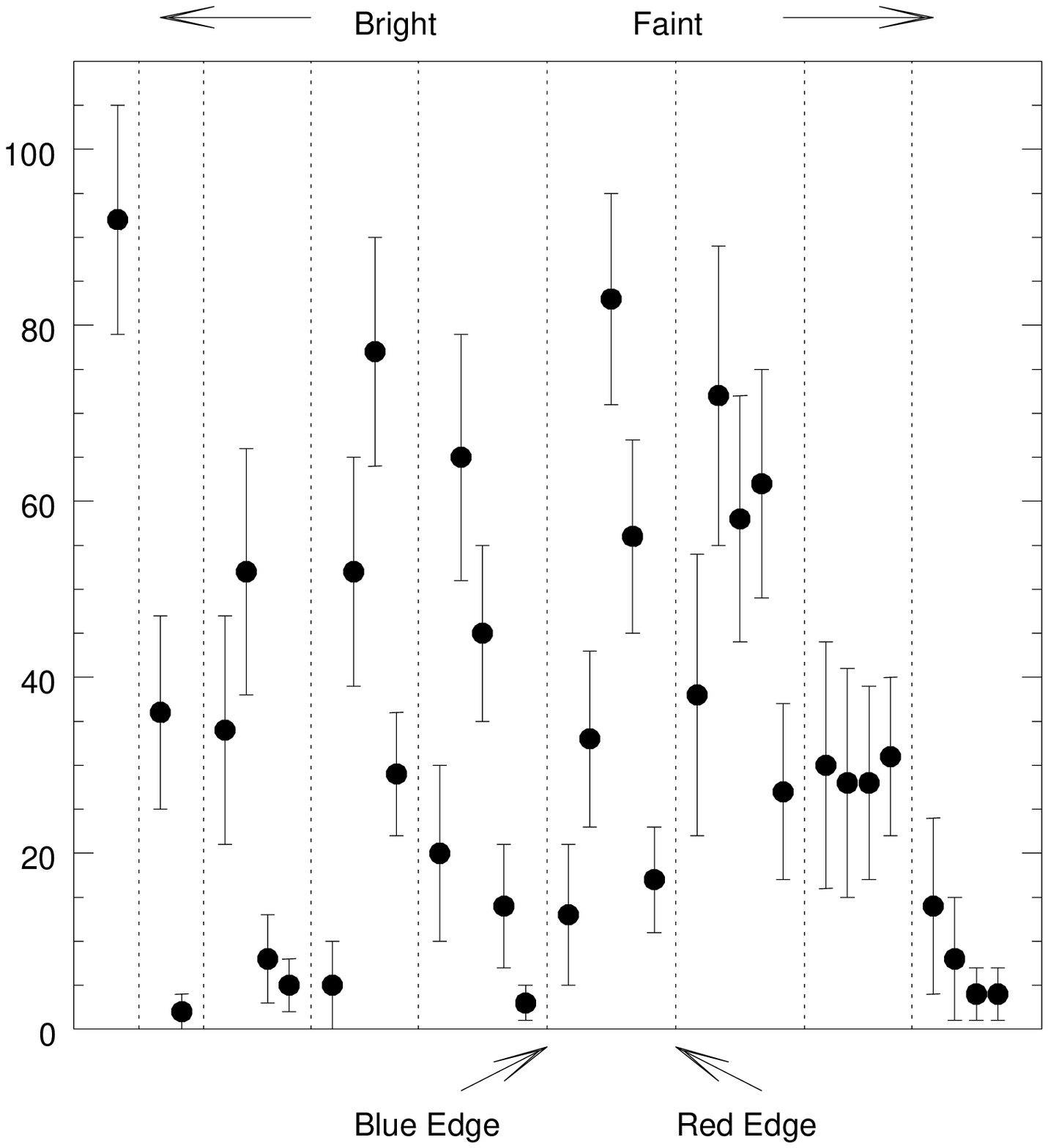}
\figcaption[Hess02.ps]{The bins indicate the sources counted in the colour-magnitude bins shown
in Figure~\ref{WDgrid}. The dotted lines delineate magnitude bins, with
the magnitude increasing as one moves from the left to the right. Between a given pair of dotted
lines, the binning indicates colour variation at fixed magnitude, with the colour getting redder
from the left to the right. It is to this data that we compare our models.
\label{Hess0}}

\clearpage
\plotone{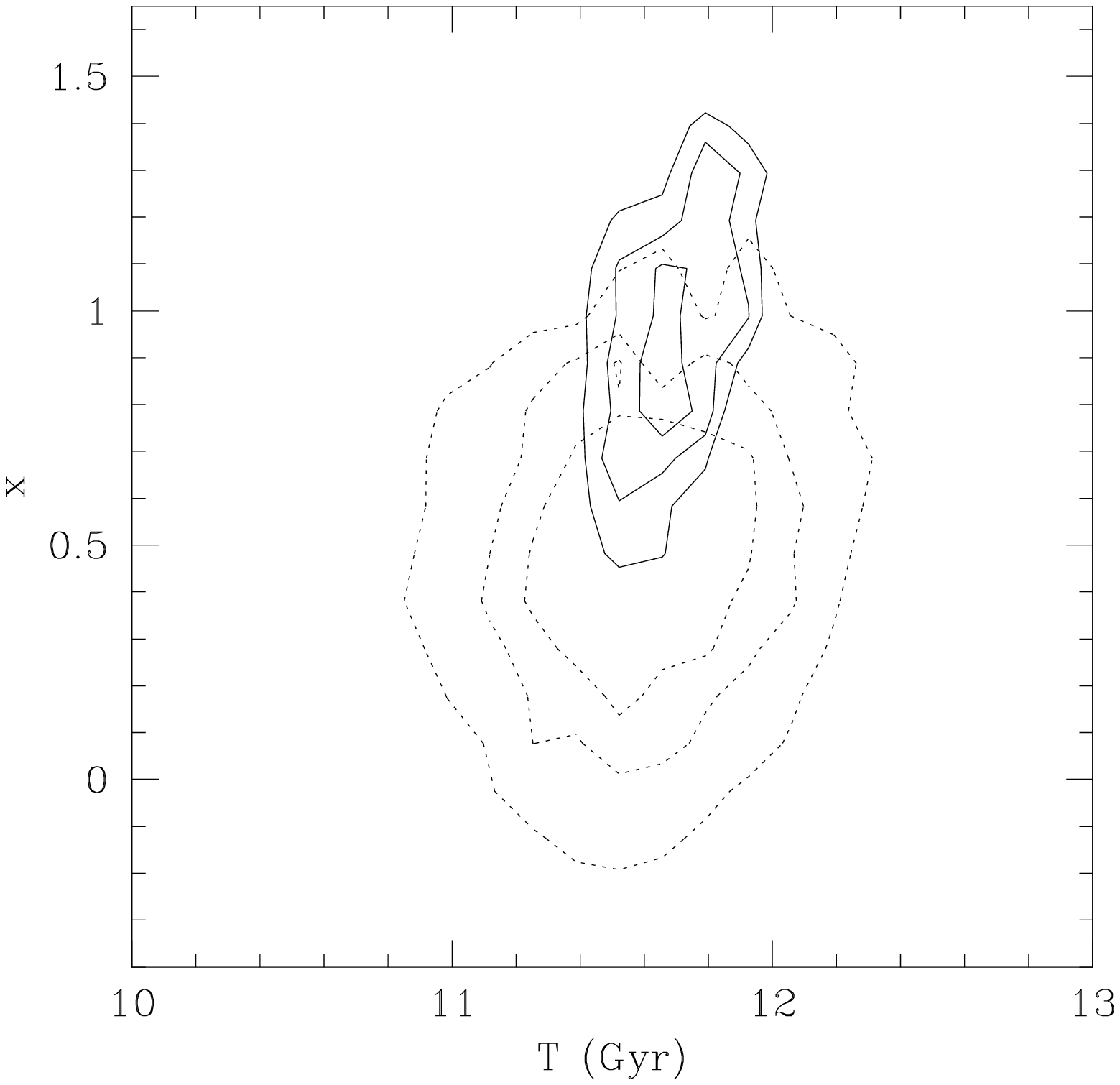}
\figcaption[xT2_smooth.ps]{The solid contours are the 1, 2 and 3 $\sigma$ contours
for the Hess diagram fit to the model. The dotted contours correspond to the
same confidence intervals, but now with the fit applied to the F814W luminosity function.
\label{xT2}}

\clearpage
\plotone{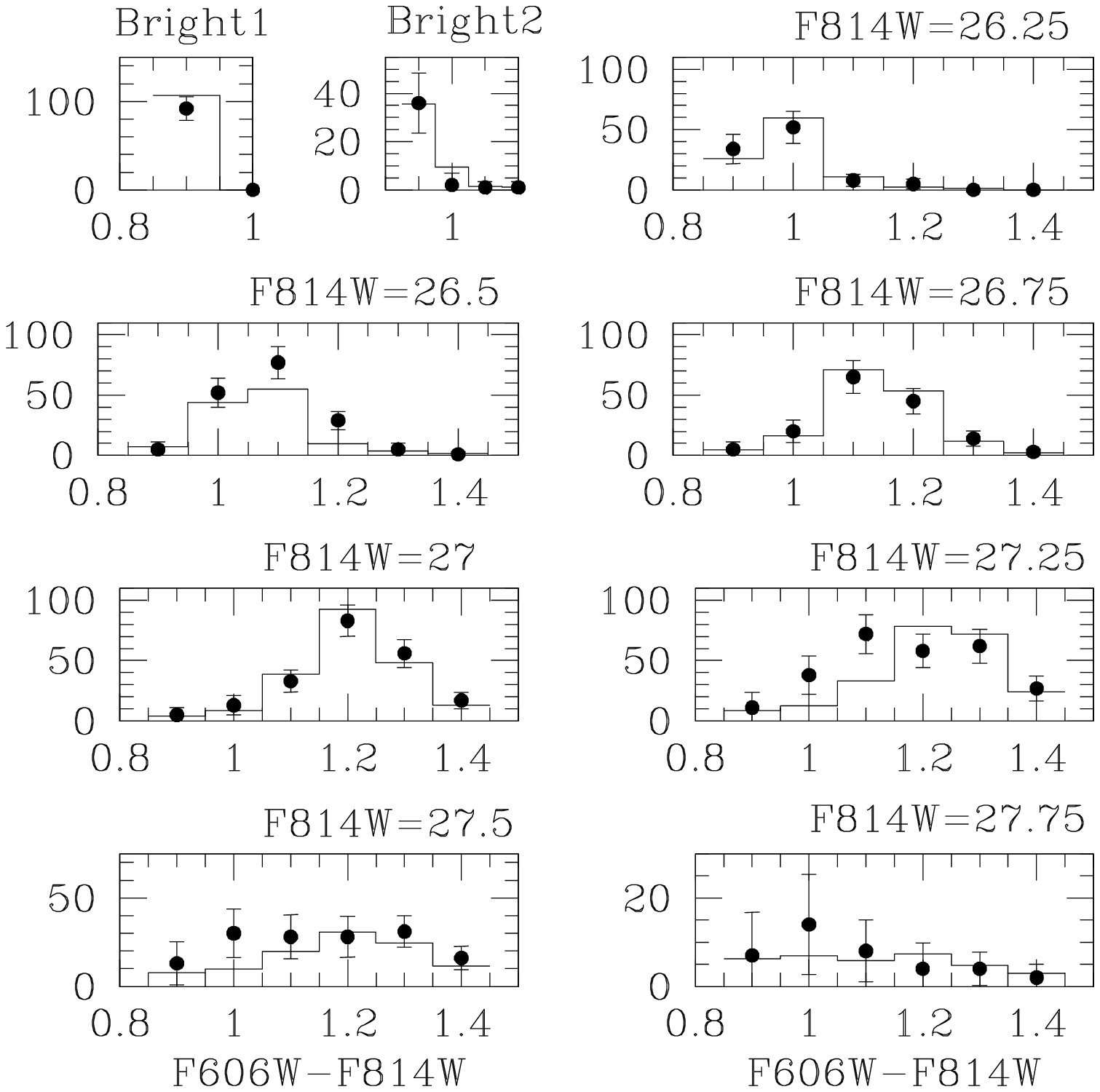}
\figcaption[Hessfit2.ps]{This model is an excellent fit overall.
The models follow the observations both in colour trends at fixed magnitude
as well as relative amplitudes as a function of magnitude. The one real discrepancy
that remains is that the the model colour distribution at F814W$=27.25$ is too red.
\label{Hessfit2}}

\clearpage
\plotone{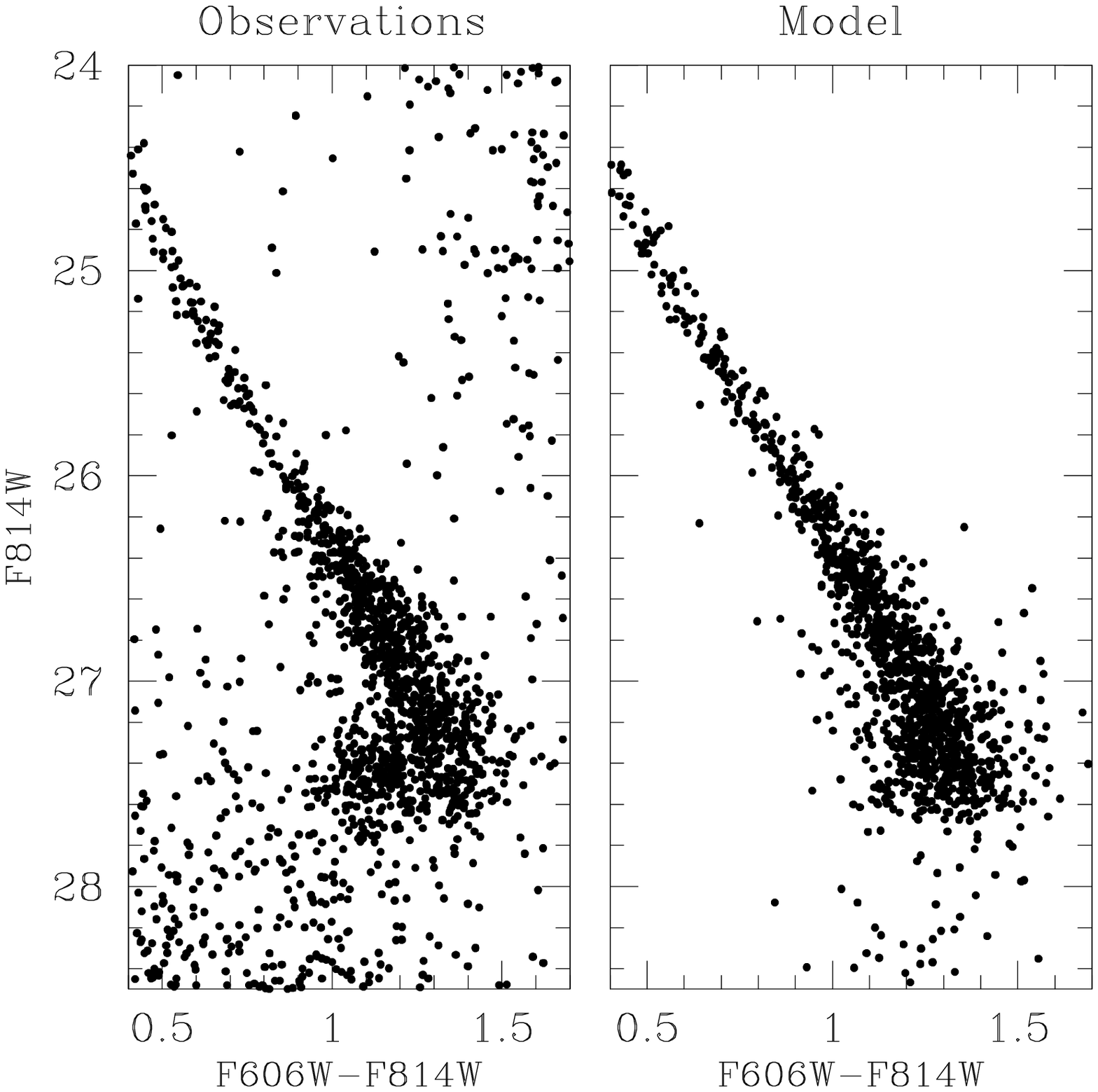}
\figcaption[Fit3.ps]{The left-hand panel shows the observed cooling sequence, after
removal of the majority of background galaxies using {\tt CENXS} cuts. The
the right-hand panel shows a monte-carlo realisation of the simulated cooling sequence,
adopting the best-fit age of 11.51~Gyr, and modeling the photometric scatter in accordance
with the artificial star tests.
\label{Fit3}}

\clearpage
\plotone{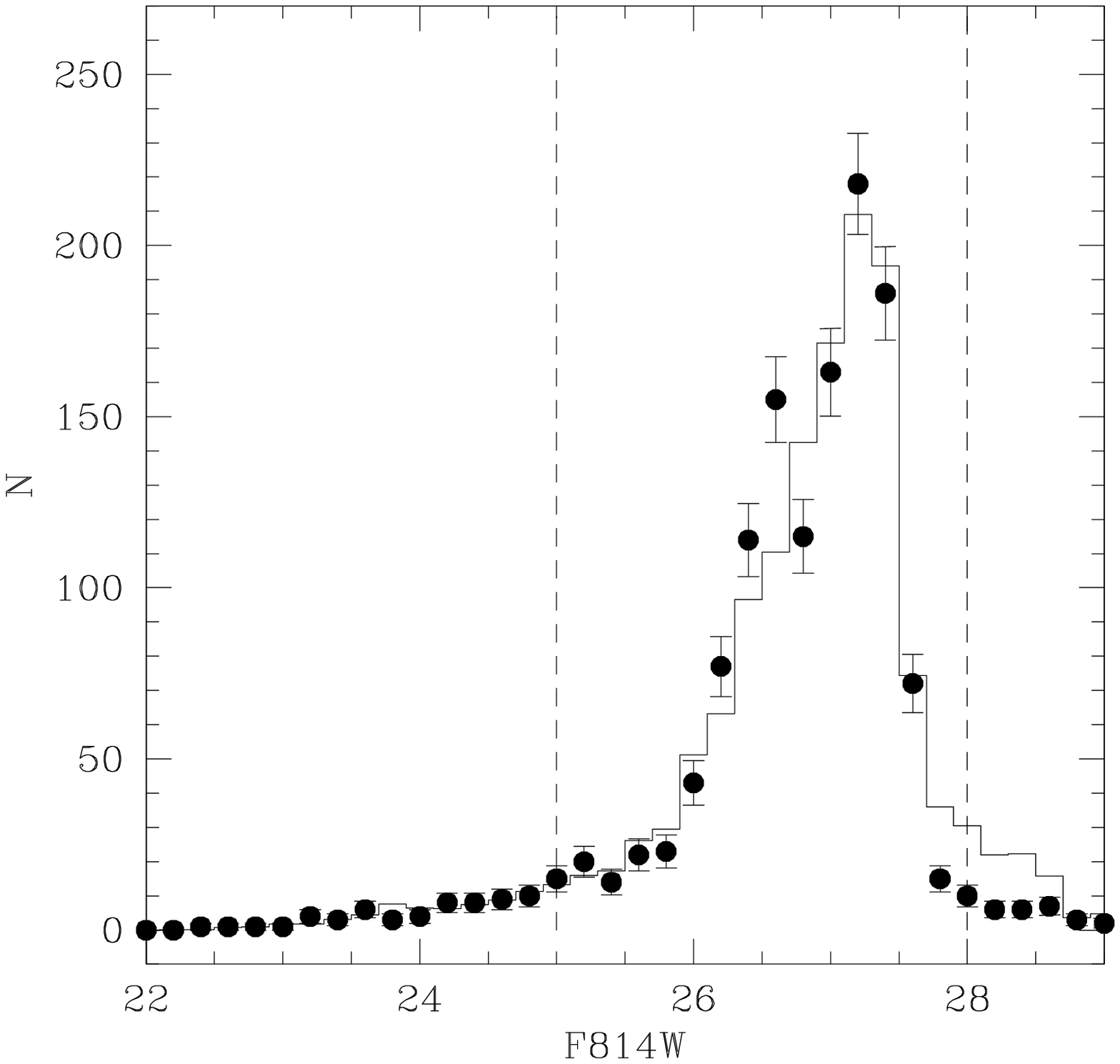}
\figcaption[LF2.ps]{The solid points are the observed luminosity function. The solid histogram
shows the model population, which includes both model white dwarfs and our estimate of the
residual galaxy contamination per bin. The
fitting region is delineated by the vertical dashed lines.
 \label{LF2}}

\clearpage
\plotone{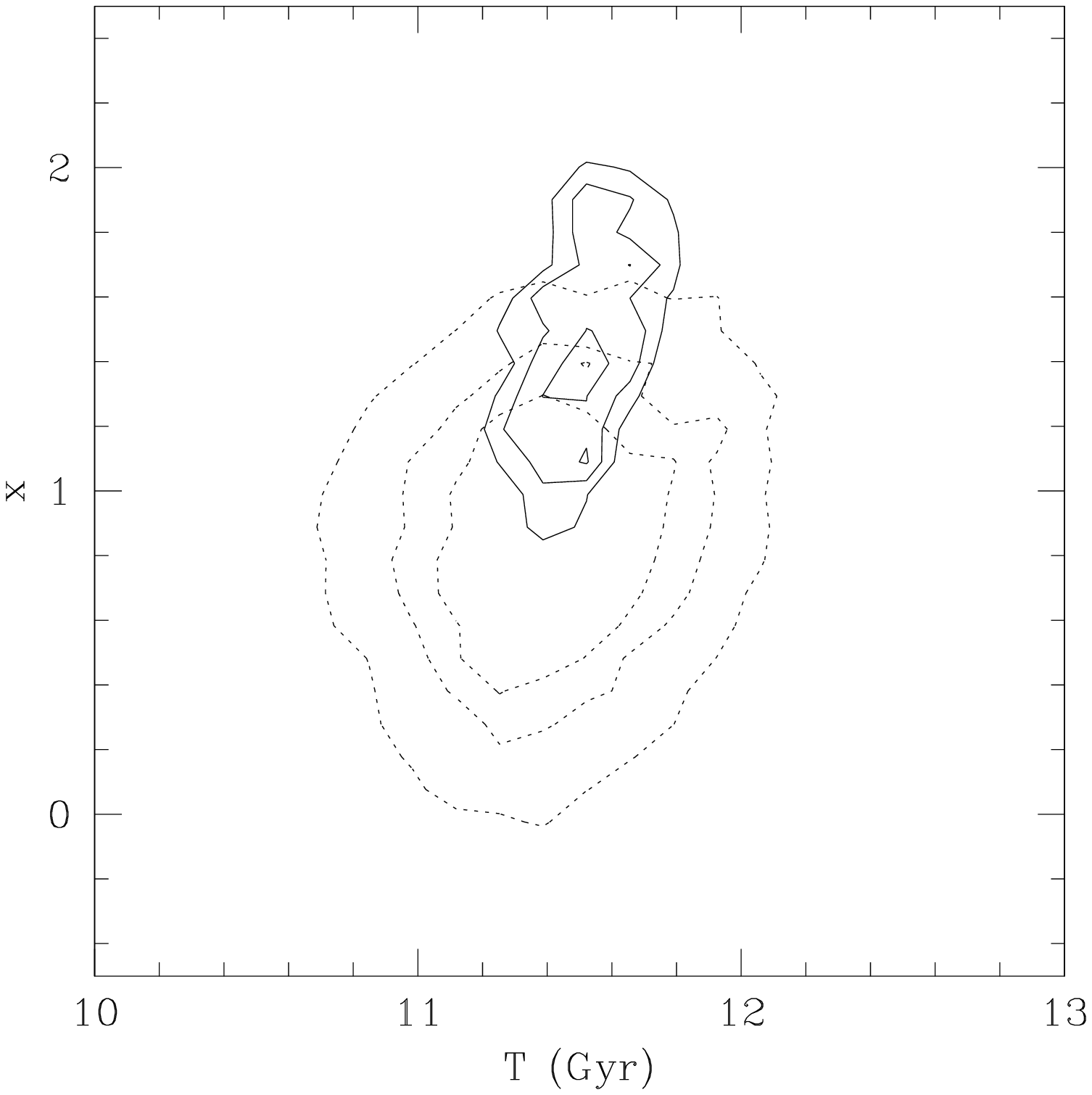}
\figcaption[xT4_smooth.ps]{The solid contours are the confidence intervals for the Hess diagram fit
but now using the $[Fe/H]=-2$ models from Dotter \& Chaboyer. The dotted contours indicate
the same but using the luminosity function.\label{xT4}}

\clearpage
\plotone{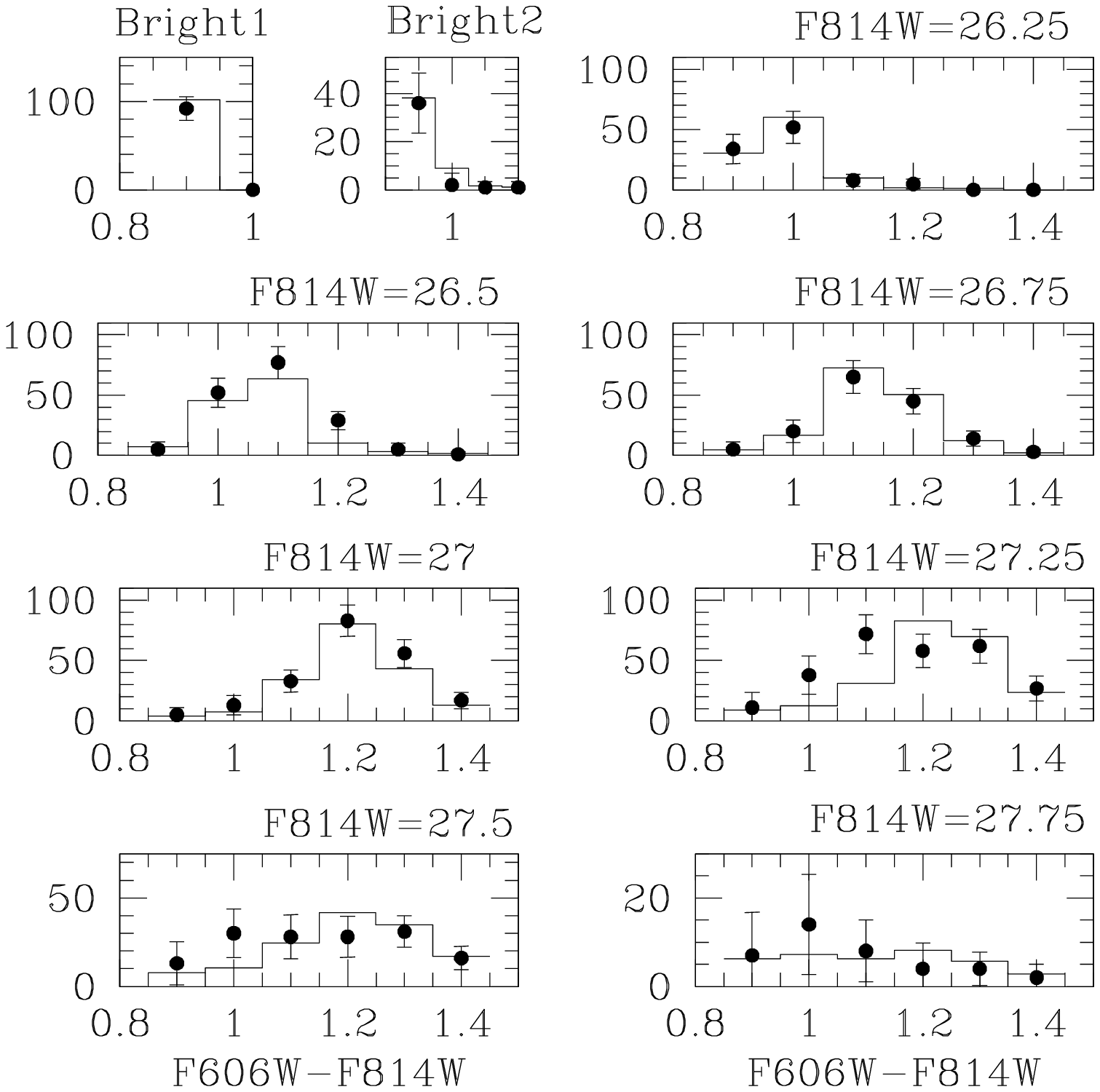}
\figcaption[Hessfit4.ps]{The histogram is once again the model fit to the observed
distribution of white dwarfs in colour-magnitude space (solid points), this time for
metal-poor progenitors and an age of 11.51~Gyr. \label{Hessfit4}}

\clearpage
\plotone{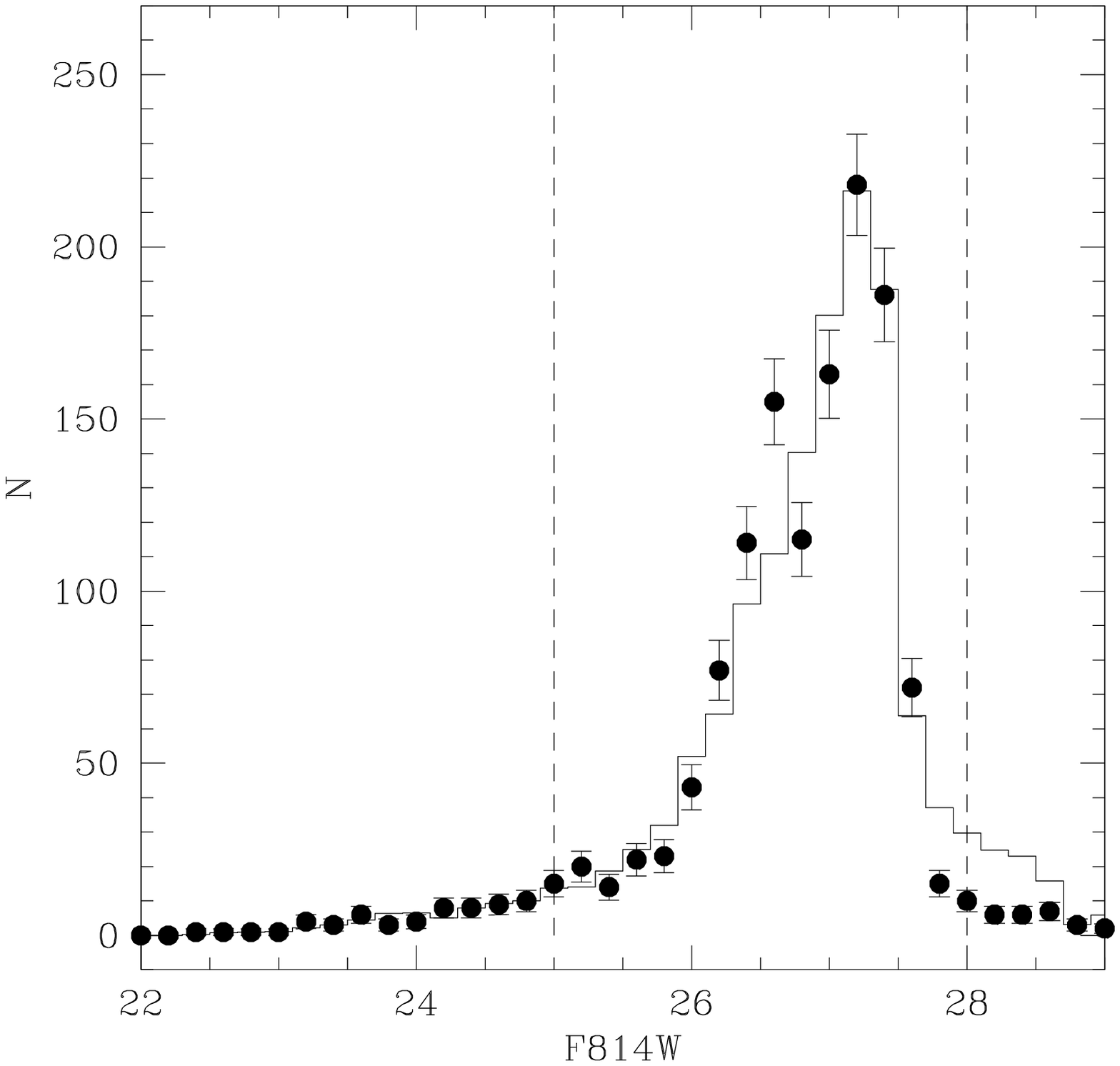}
\figcaption[LF4.ps]{The model luminosity function shown is for an age of 11.46~Gyr.\label{LF4}}

\clearpage
\plotone{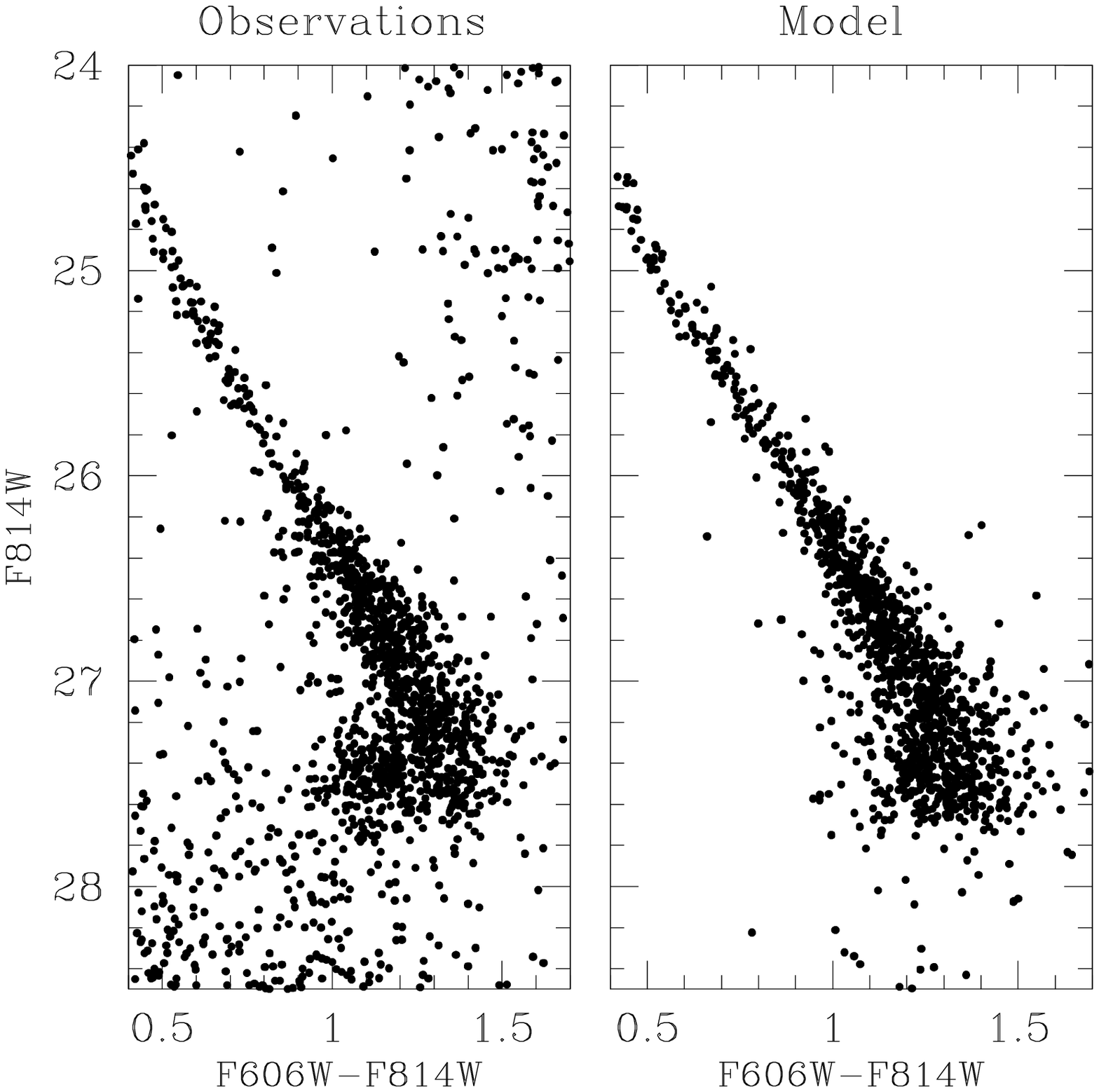}
\figcaption[Fit3_4.ps]{The left-hand panel is the observed cooling sequence while the right-hand
side is a realisation of an 11.51~Gyr old model with metal-poor progenitors. \label{Fit3_4}}

\clearpage
\plotone{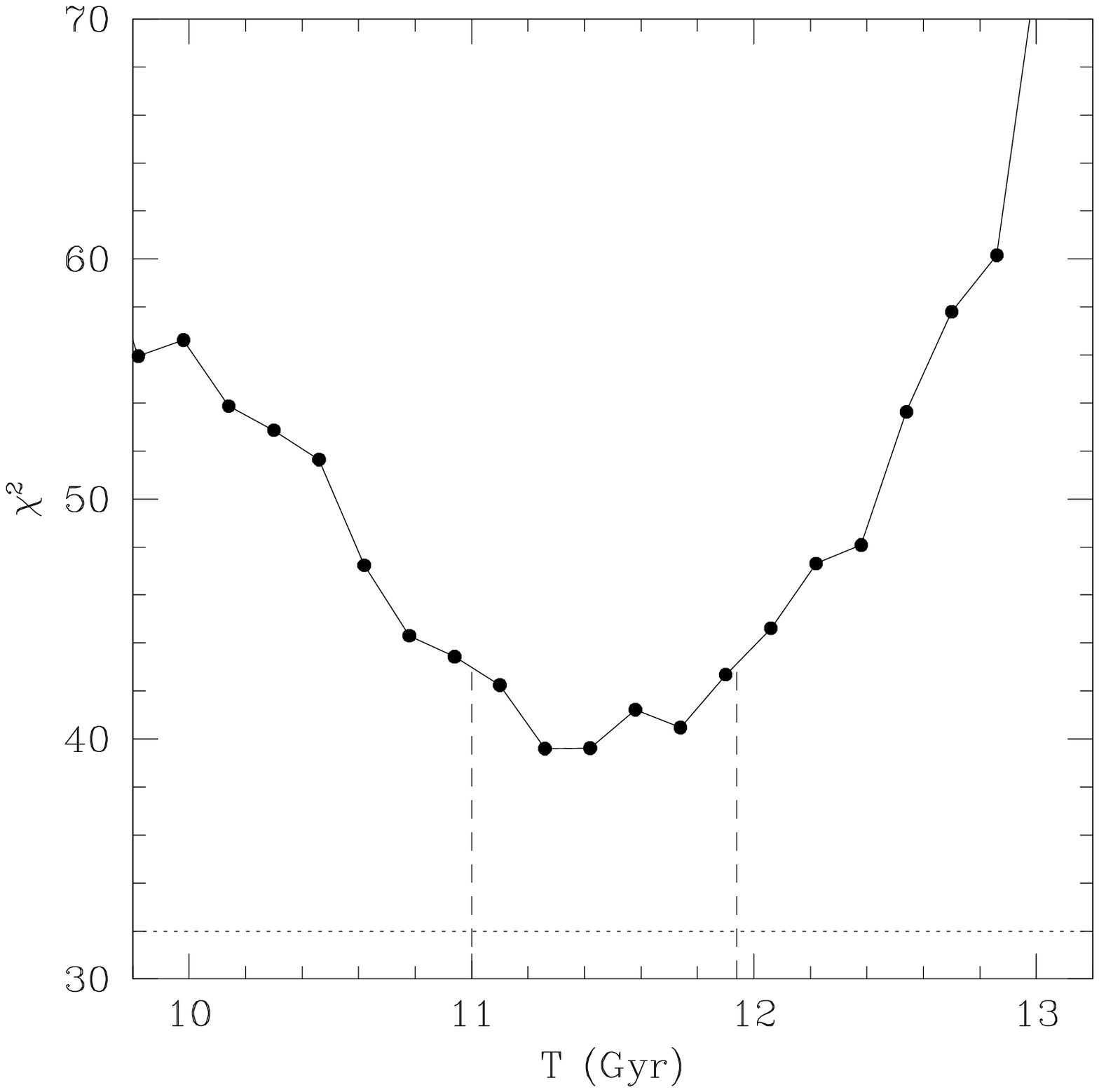}
\figcaption[Tc.ps]{The $\chi^2$ curve versus cluster age, marginalised over all other parameters for
the metal-poor main sequence models (both those of Dotter \& Chaboyer and those of Hurley et al). 
The dashed lines indicate the $2 \sigma$ age range
and the horizontal dotted line corresponds to $\chi^2=1$ per degree of freedom.
\label{Tc}}

\clearpage
\plotone{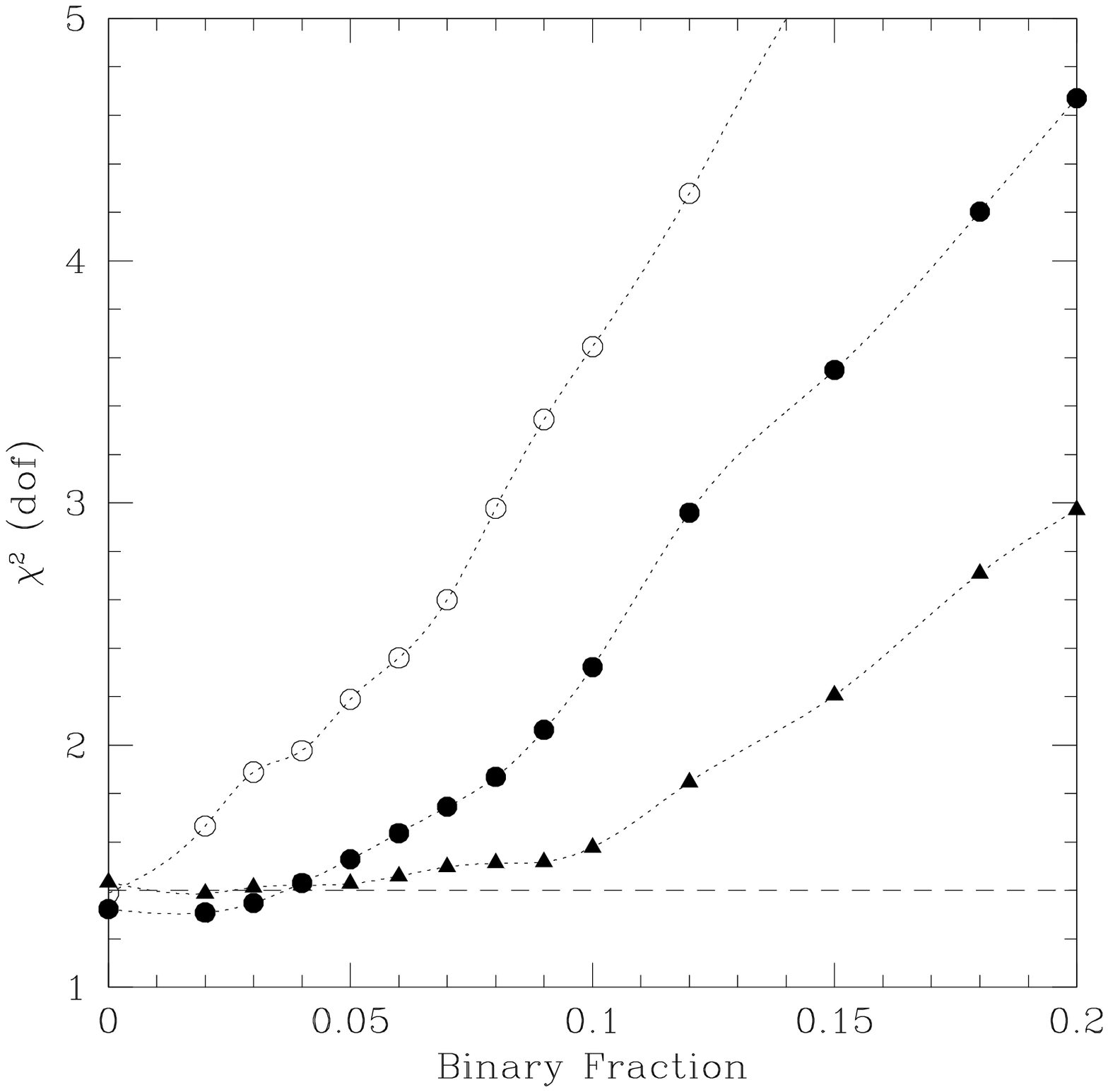}
\figcaption[Bin.ps]{The open circles are the $\chi^2$ per degree of freedom for fits to the
observed F814W luminosity function, as a function of white dwarf binary fraction. The filled
circles indicate fits to the Hess diagram - triangles in the case of our normal grid, and
circles in the case of the expanded grid. The dashed line indicates the $\chi^2$(dof) limit
on the expanded Hess fit.
\label{Bin}}

\clearpage
\plotone{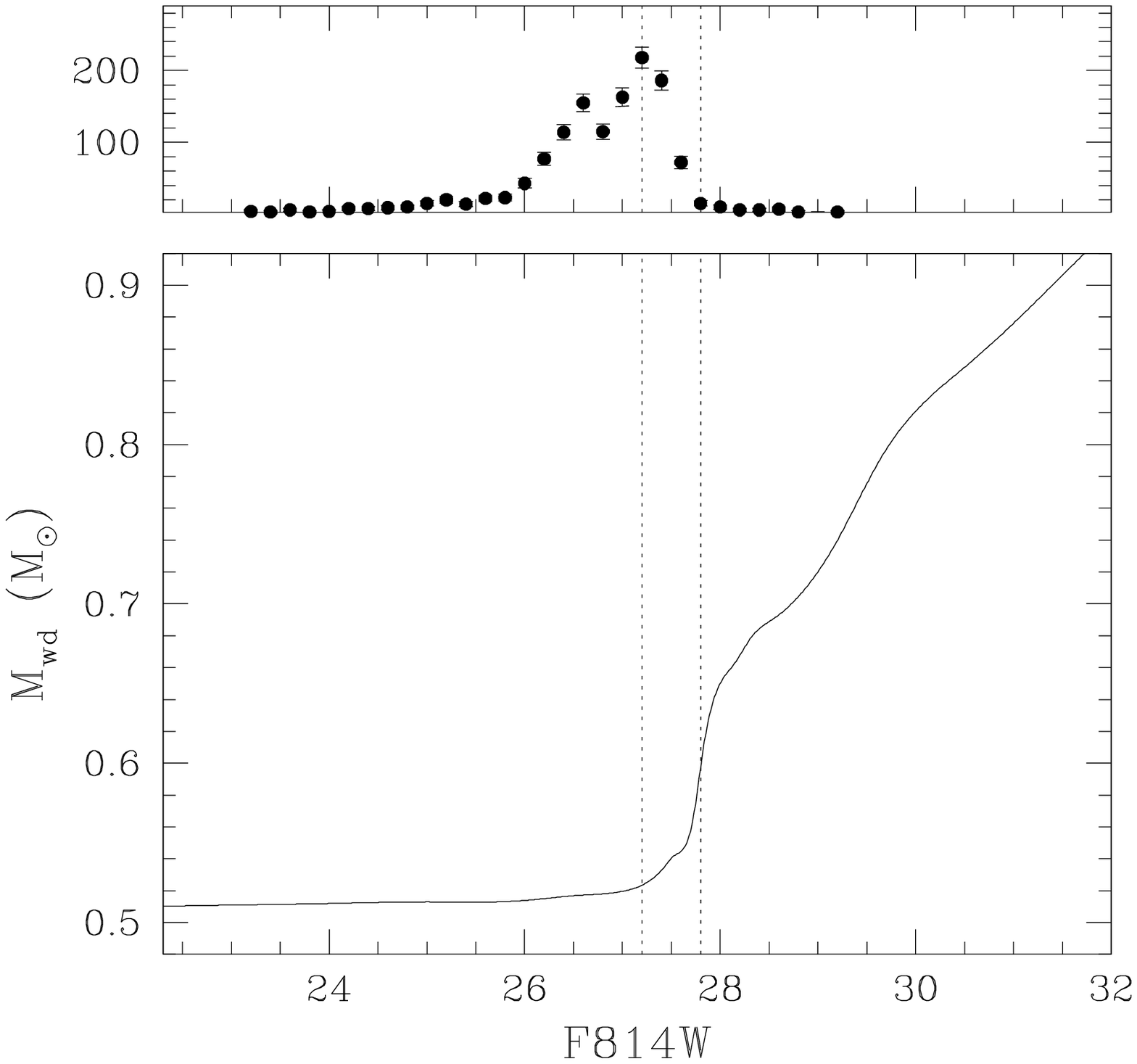}
\figcaption[MM.ps]{The upper panel shows our observed luminosity function, while the
lower panel shows the relationship between white dwarf mass and F814W magnitude in our
best-fit model. We see that the bulk of the luminosity function contains white dwarfs
of the same mass and that the model masses only start to increase significantly as we
reach the truncation of the luminosity function. In fact, it is the faster cooling of
more massive white dwarfs at late times that causes this truncation. Note that this is
at odds with the common interpretation of such a feature as being due to the faintest
white dwarfs that exist in the cluster.
\label{MM}}

\clearpage
\plotone{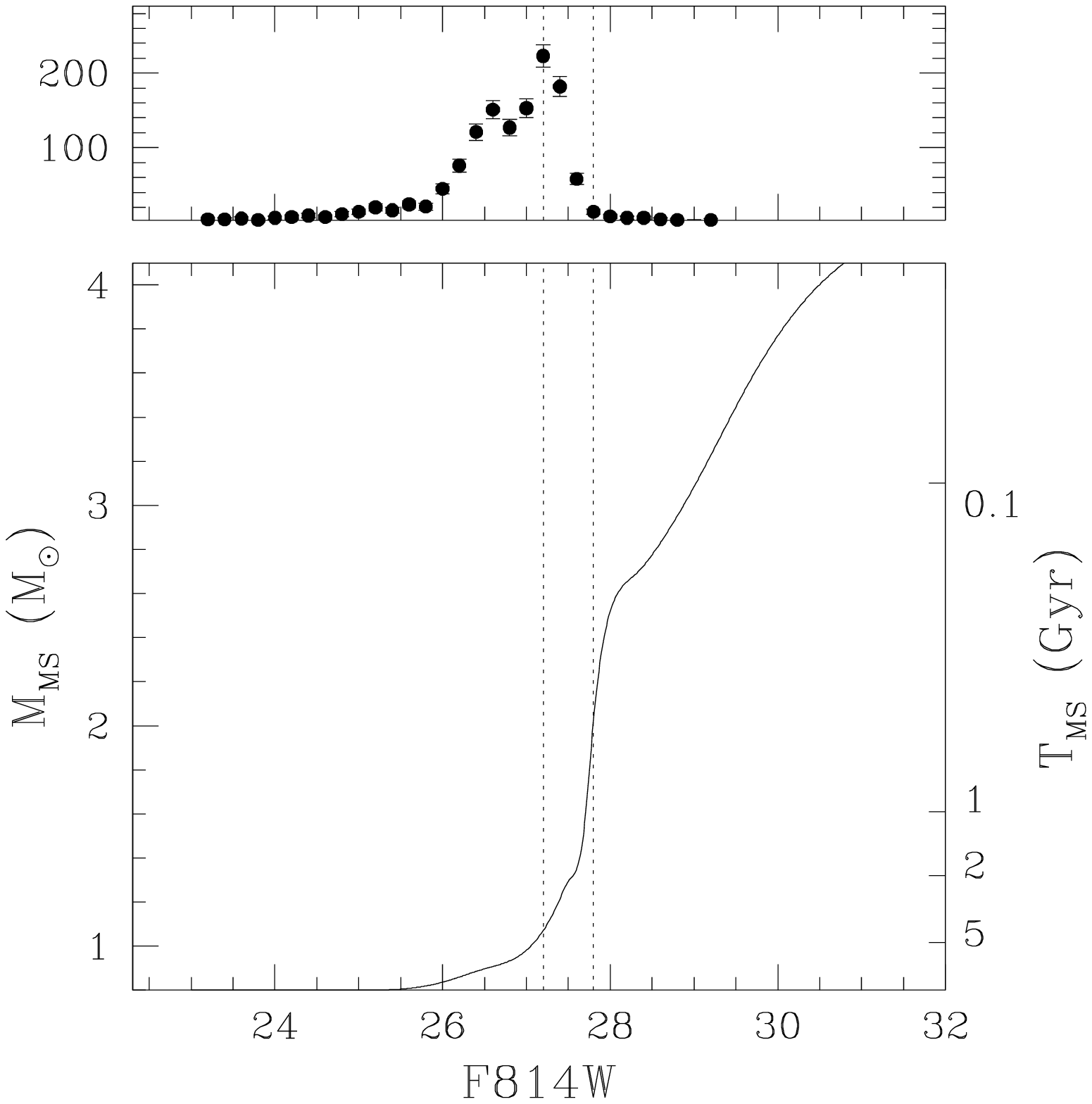}
\figcaption[MM2.ps]{This figure is similar to Figure~\ref{MM} expect we now show the
main sequence mass of the white dwarf progenitors in the best fit model. We see that
the white dwarf luminosity function probes only a rather limited range in stellar masses.
\label{MM2}}

\clearpage
\plotone{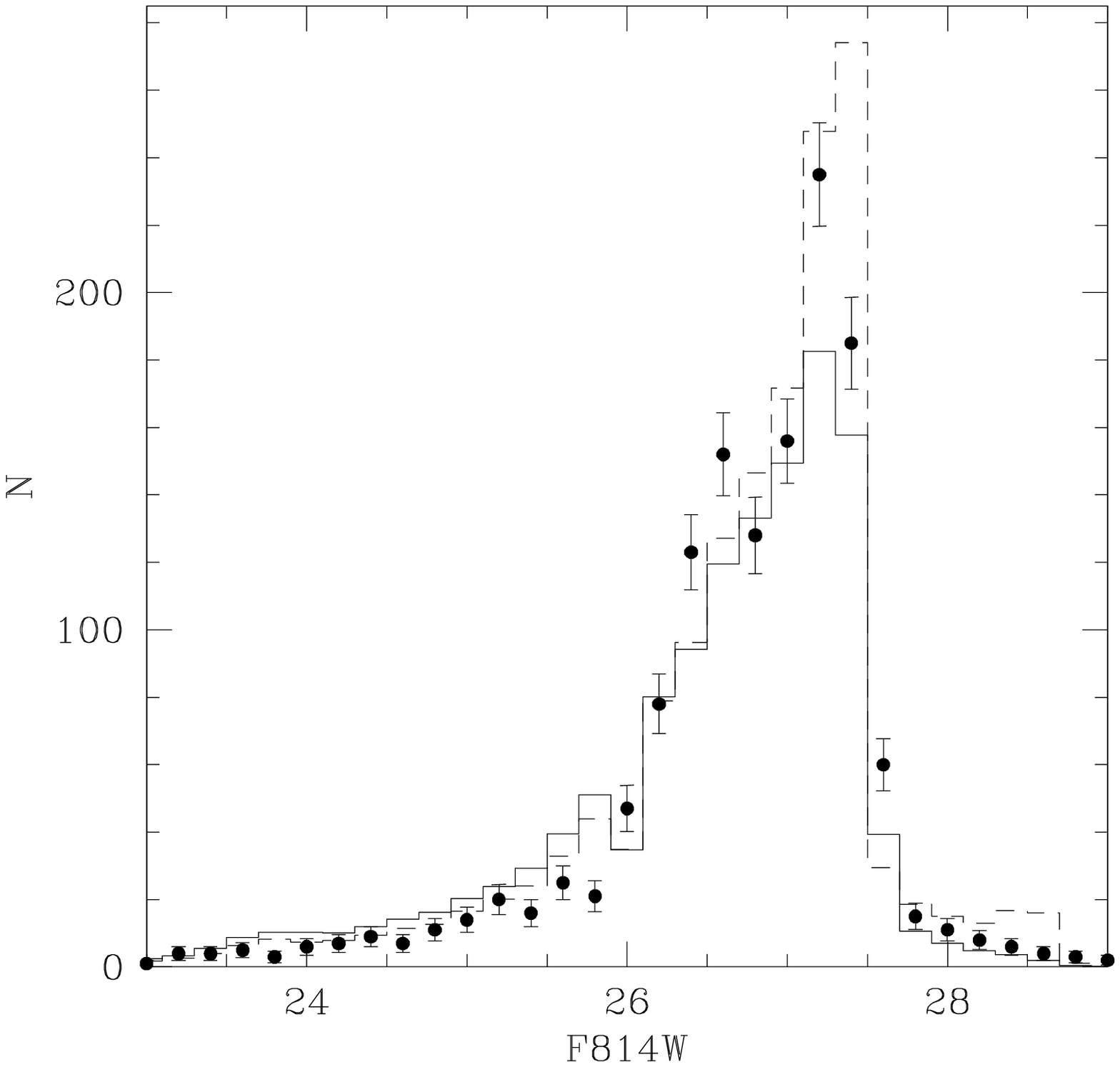}
\figcaption[IncOff.ps]{This figure demonstrates that the luminosity function truncation
is a real feature of the white dwarf population and not simply a function of observational
incompleteness. The solid histogram is the best-fit model to the observed luminosity function
and the dashed histogram is the very same model but now without taking into account the loss
of stars due to incompleteness. We see that the sharp peak is slightly smoothed out by the
loss of stars due to crowding, but that the sharp truncation is still preserved.
\label{IncOff}}

\clearpage
\plotone{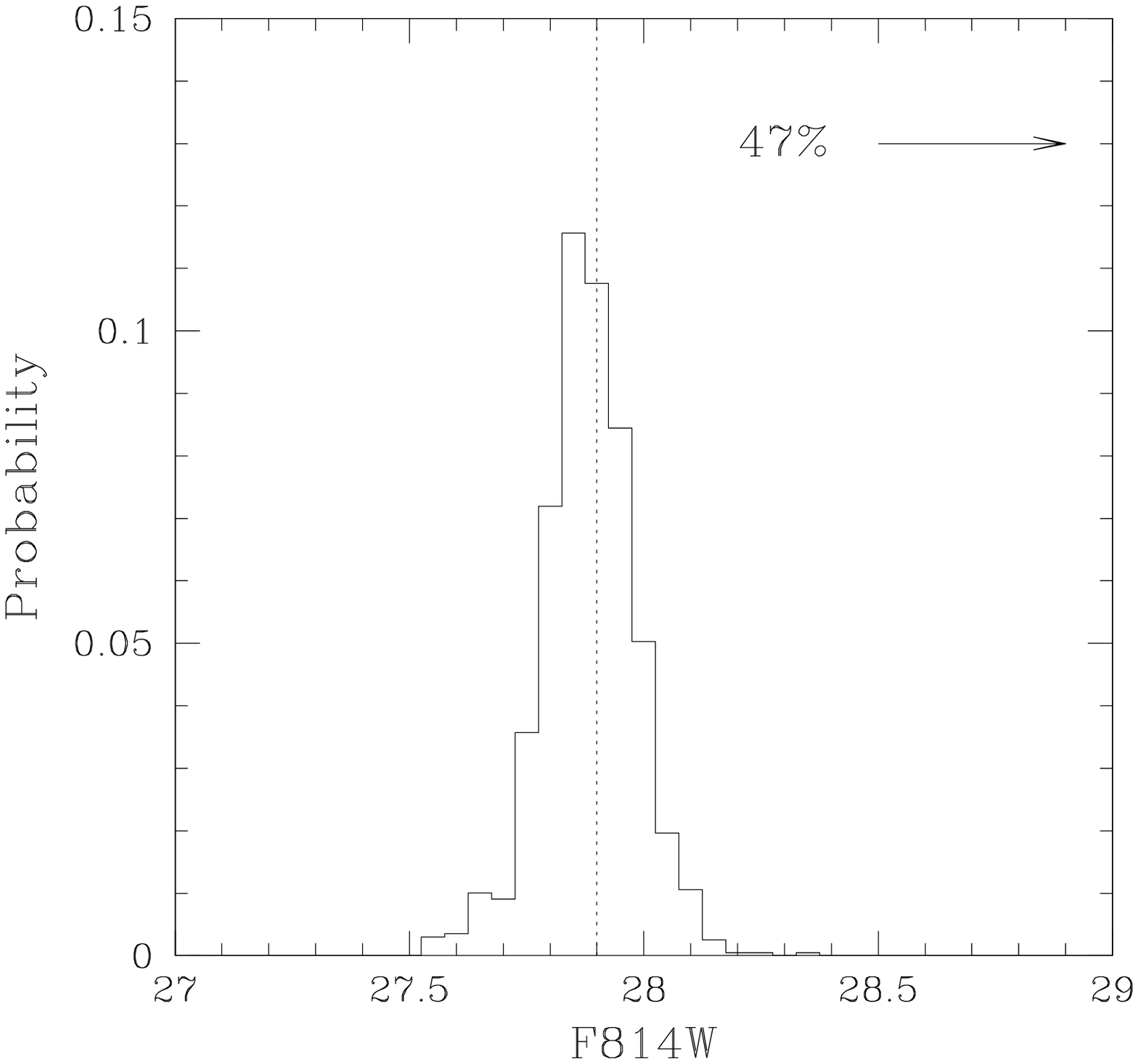}
\figcaption[Spread.ps]{This figure illustrates the nature of our observational incompleteness,
as determined by artificial star tests. The dotted line indicates the input magnitude (chosen
to represent the end of the luminosity function) and the
solid histogram indicates the spread in magnitude of those artificial stars successfully recovered.
Note that this population represents only 53\% of the input stars. The others are never recovered
because they are located too close to much brighter stars.
\label{Spread}}

\clearpage
\plotone{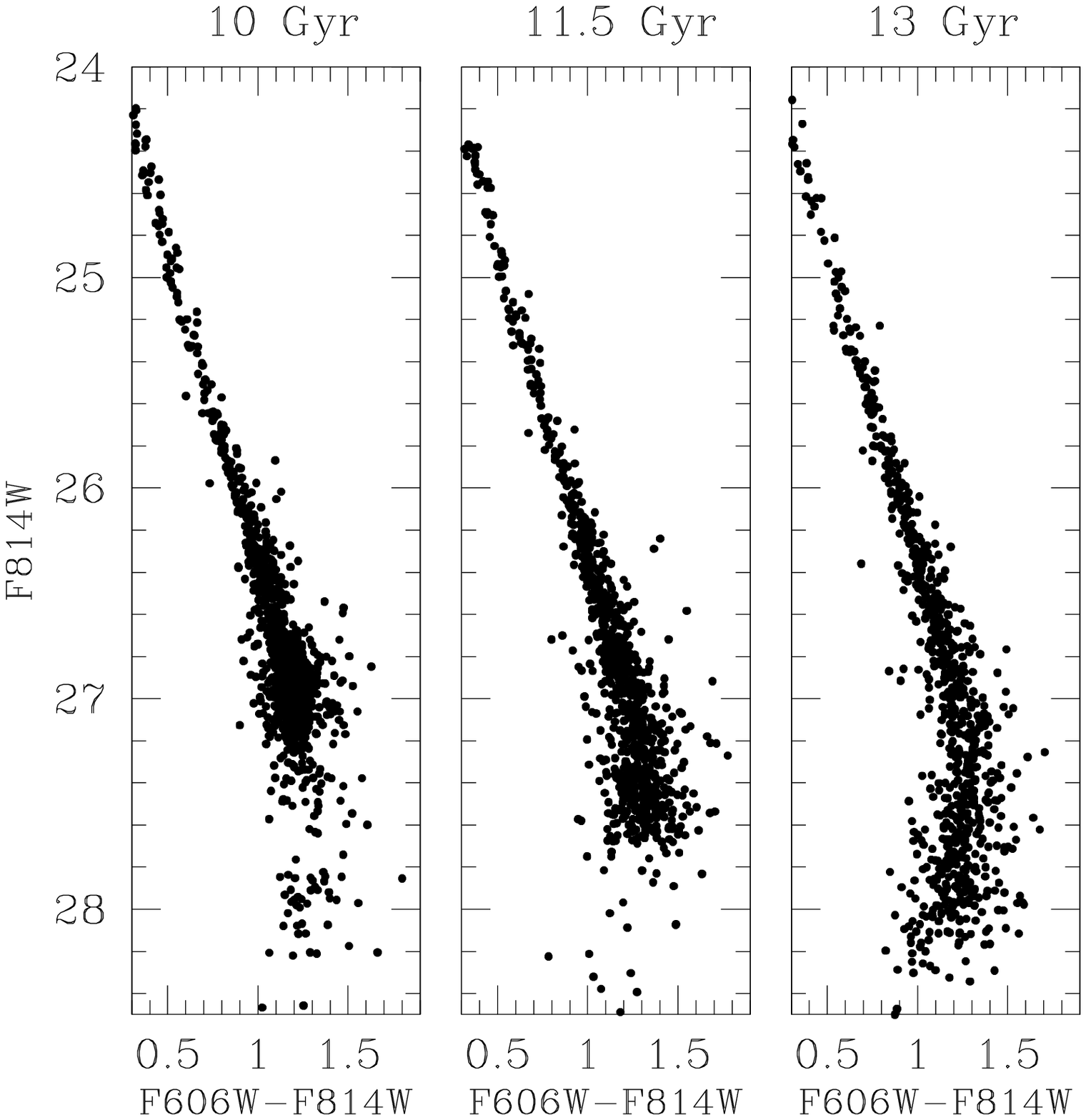}
\figcaption[Fit4.ps]{The models in all three panels are identical except for the assumed age.
The center panel has an age of 11.5~Gyr and corresponds to our best-fit model for NGC~6397.
The left hand panel shows a younger population, with age of 10~Gyr, and the right hand panel
shows an older one, with age 13~Gyr. It is clear from a simple visual inspection
 that neither of the side panels are a good match to the observed population seen in 
Figure~\ref{Fit3} or \ref{Fit3_4}.
\label{Fit4}}

\clearpage
\plotone{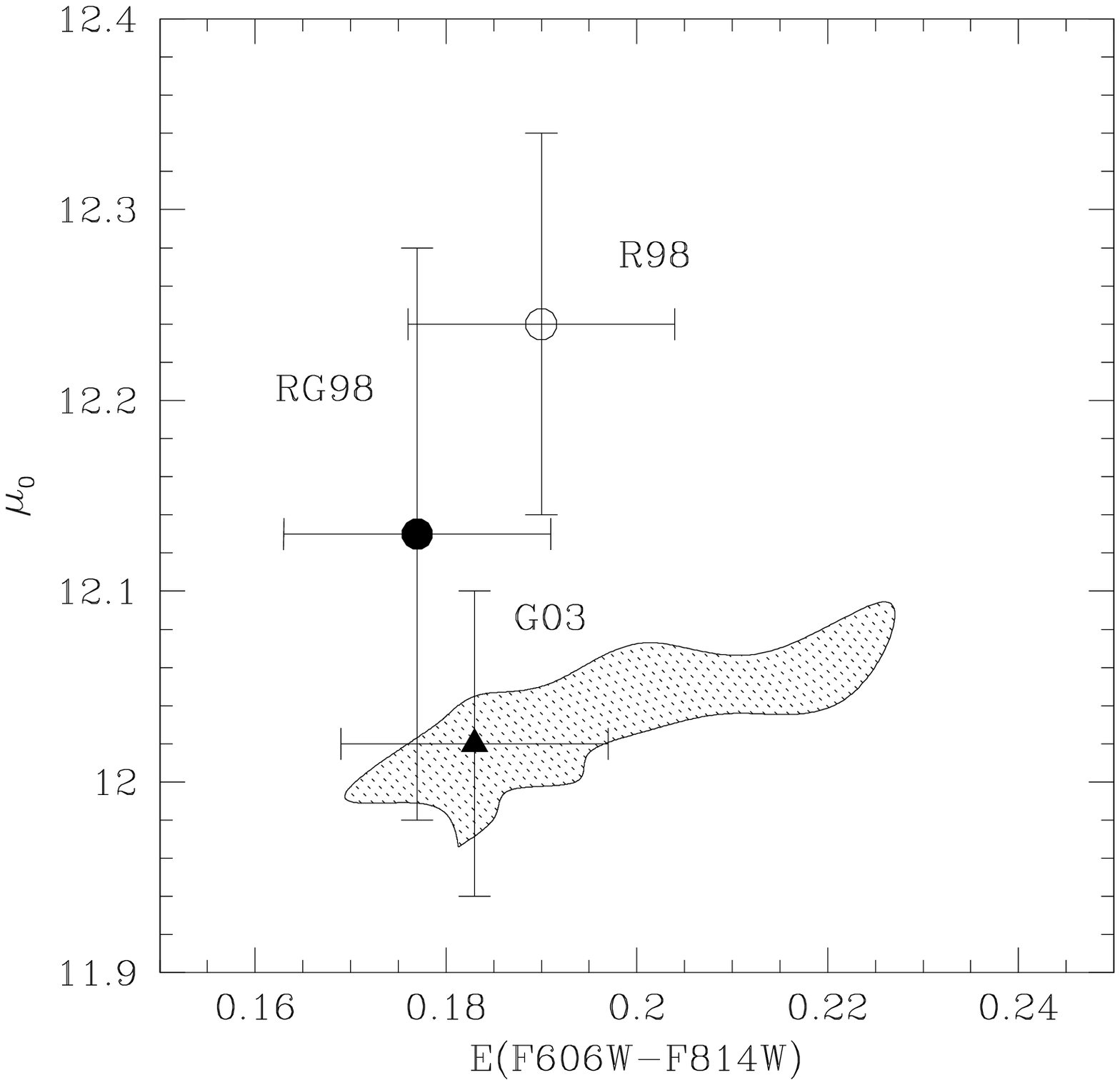}
\figcaption[MuA.ps]{The shaded region is the 2$\sigma$ age range for distance and reddening allowed by fitting
to the white dwarf cooling sequence. The solid point with the error bar represents the
distance inferred by Reid \& Gizis (1998), plotted at their assumed reddening. Similarly, the inferred
values from Reid (1998) and Gratton et al (2003) are also shown. The reddenings have been converted from the ground-based colours to the ACS photometric system using the prescription of Sirianni et al (2005). The horizontal error bars represent the effect of assuming different underlying spectral types, spanning M5 to O2, when
converting from ground-based to HST magnitudes.
\label{MuA}}

\clearpage
\plotone{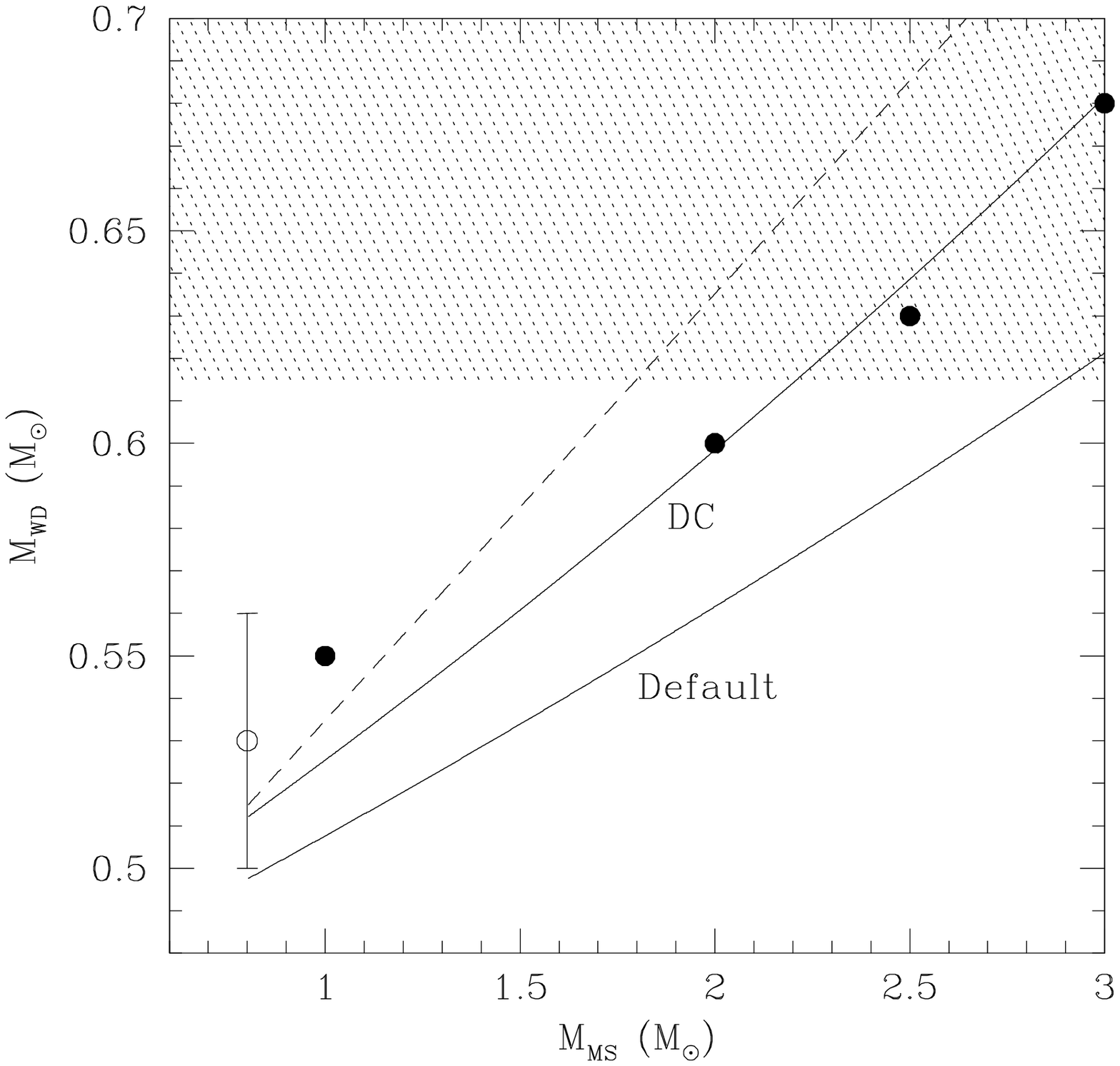}
\figcaption[IFMR.ps]{The solid lines indicate our best fit initial-final mass relations
for the default models and for the metal-poor models with Dotter \& Chaboyer main sequence
lifetimes.  The solid points are the empirical relation inferred by Weidemann (2000) and
the dashed line indicates the best-fit linear relation espoused by Ferrario et al (2005).
The open circle, with the error bar, is the mean upper cooling sequence mass inferred by
Moehler et al (2004).
Our fit is only constrained for a limited range of white dwarf masses (see Figure~\ref{MM})
and so we shade the range of white dwarf masses that fall beyond our magnitude limit.
We see that our best-fit models overall (the DC models) are in excellent agreement with
Weidemann's empirical determination.
\label{IFMR}}

\clearpage
\plotone{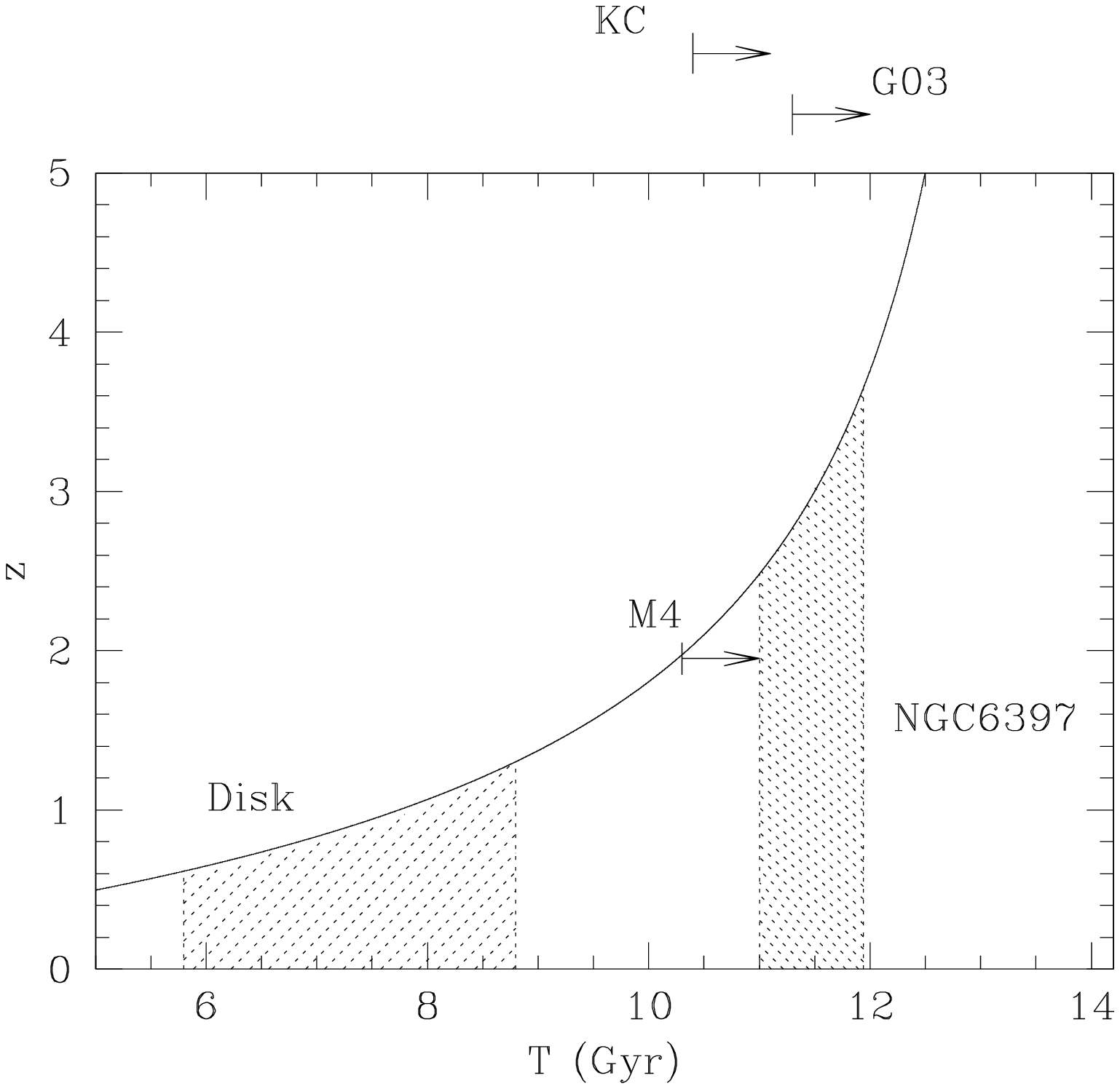}
\figcaption[Age.ps]{The solid curve indicates the relationship between cosmological redshift $z$
and lookback time for the best-fit flat universe model from Spergel et al (2003).  The shaded
regions indicate white dwarf cooling ages (2$\sigma$ range) for the Galactic disk (Hansen et al 2002) and
NGC~6397 (this paper) and the arrow indicates the lower limit on the age for M4 (Hansen et al 2004).
Above the plot we show two 95\% lower limits. The limit marked KC indicates the lower limit for the age
of the globular cluster system as whole, taken from Krauss \& Chaboyer (2003). The limit marked G03 is
the 2$\sigma$ lower limit on the age of NGC6397, based on the results of Gratton et al (2003). The comparison
indicates that our age determination is consistent with, but also more accurate than, the best measurements
using the MSTO method.
\label{Age}}

\clearpage
\plotone{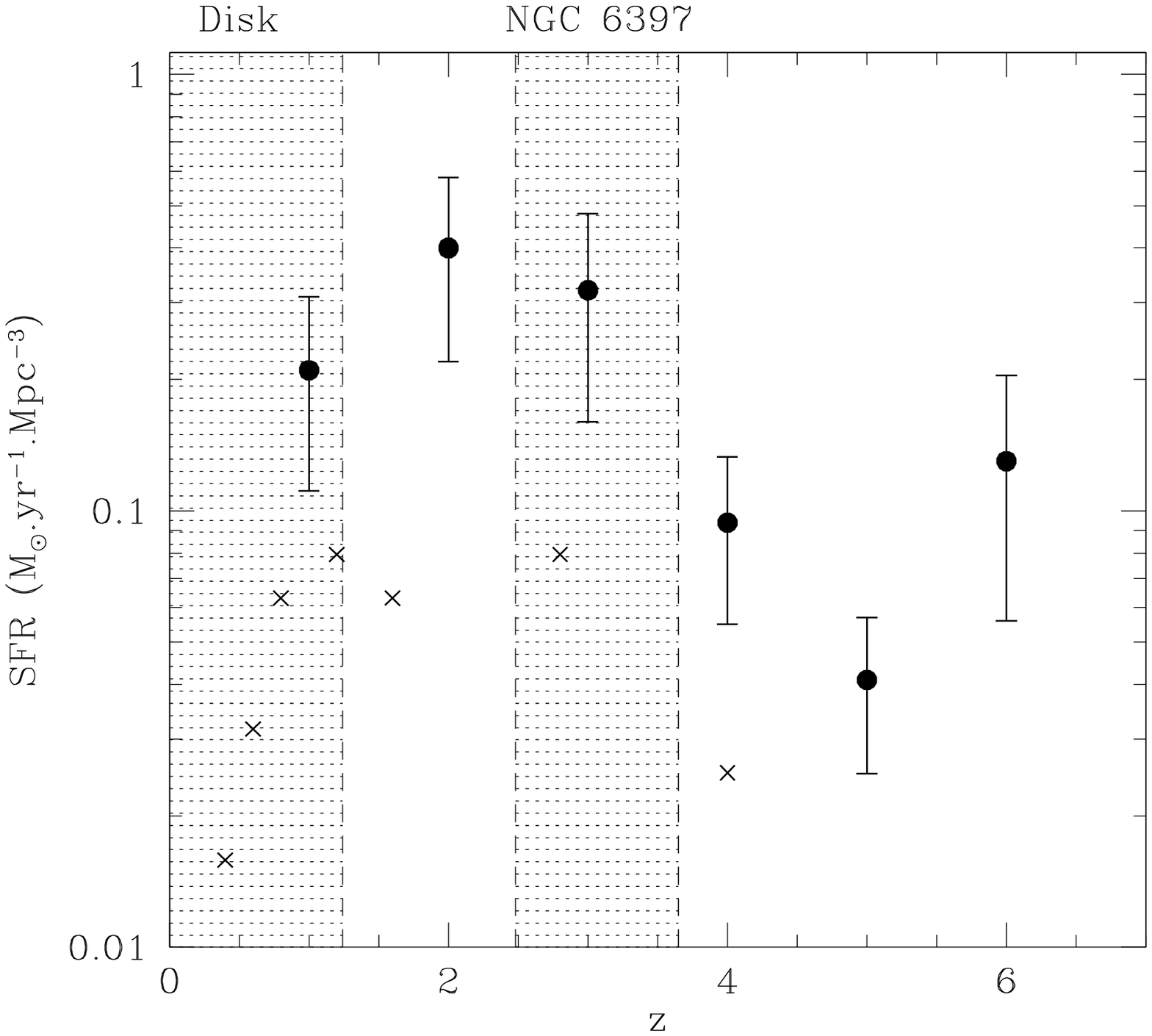}
\figcaption[SFR.ps]{The solid points represent the Hubble Ultra-Deep Field star formation
rates (corrected for extinction and surface brightness) measured by Thompson et al (2006).
The crosses are the extinction corrected values measured from the Hubble Deep Field by
Madau et al (1996). The shaded regions indicate the epoch in which the Galactic Disk and
NGC~6397 formed, according to our measures of the white dwarf cooling.
\label{SFR}}

\clearpage
\plotone{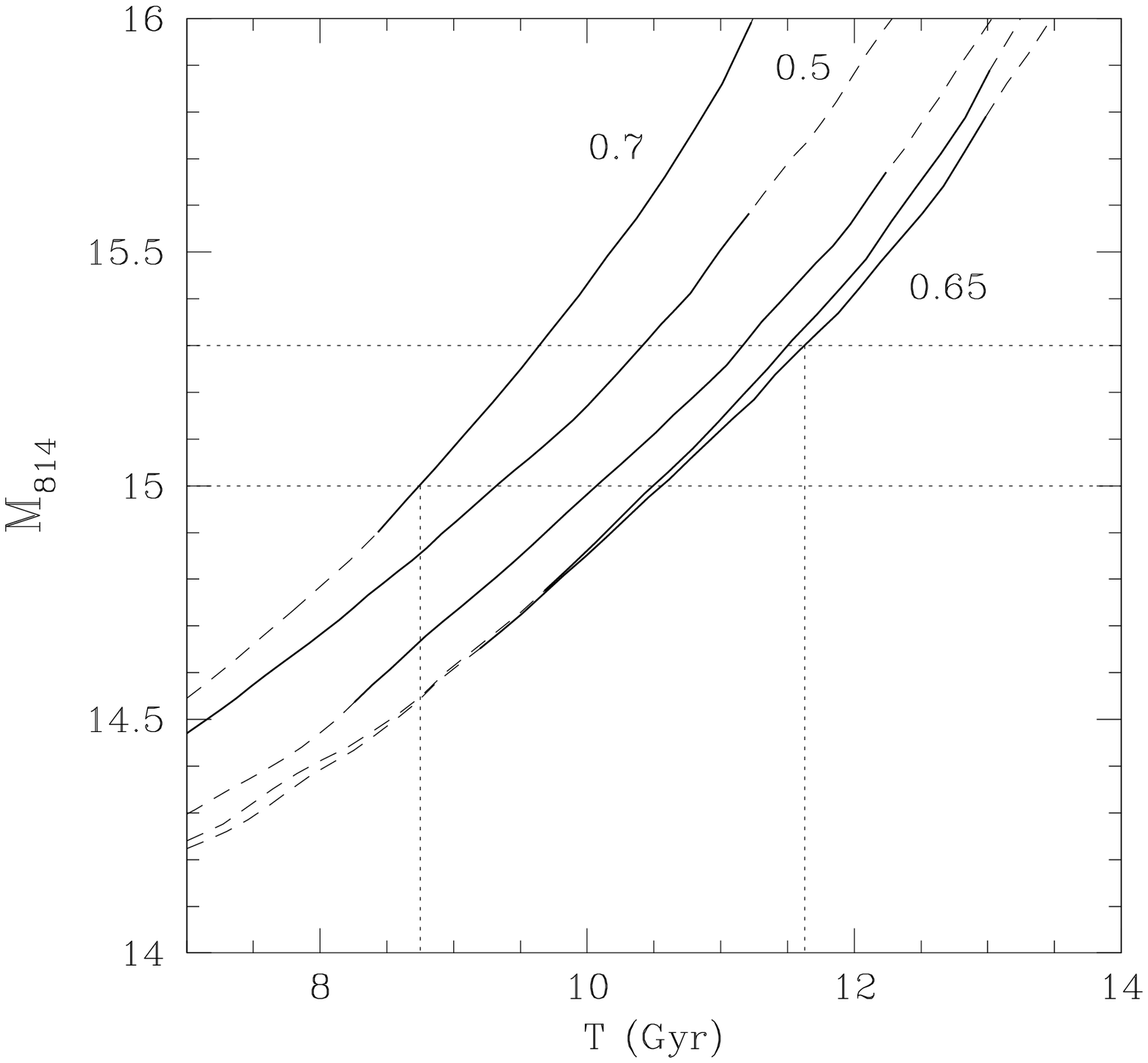}
\figcaption[Curve.ps]{The horizontal dotted lines indicate the inferred absolute magnitude range where the truncation
in the cooling sequence occurs. The various curves show the results for cooling models of different masses (except
for the leftmost curve, labelled $0.7 M_{\odot}$, the mass increases from left to right, in increments of $0.05 M_{\odot}$, from that labelled $0.5 M_{\odot}$ to that labelled $0.65 M_{\odot}$). Those parts of each curve that are solid correspond to F606W-F814W colours consistent with those at the observed truncation, while the dashed parts do not.
The vertical dotted lines indicate the inferred age range.
\label{Curve}}

\clearpage
\plotone{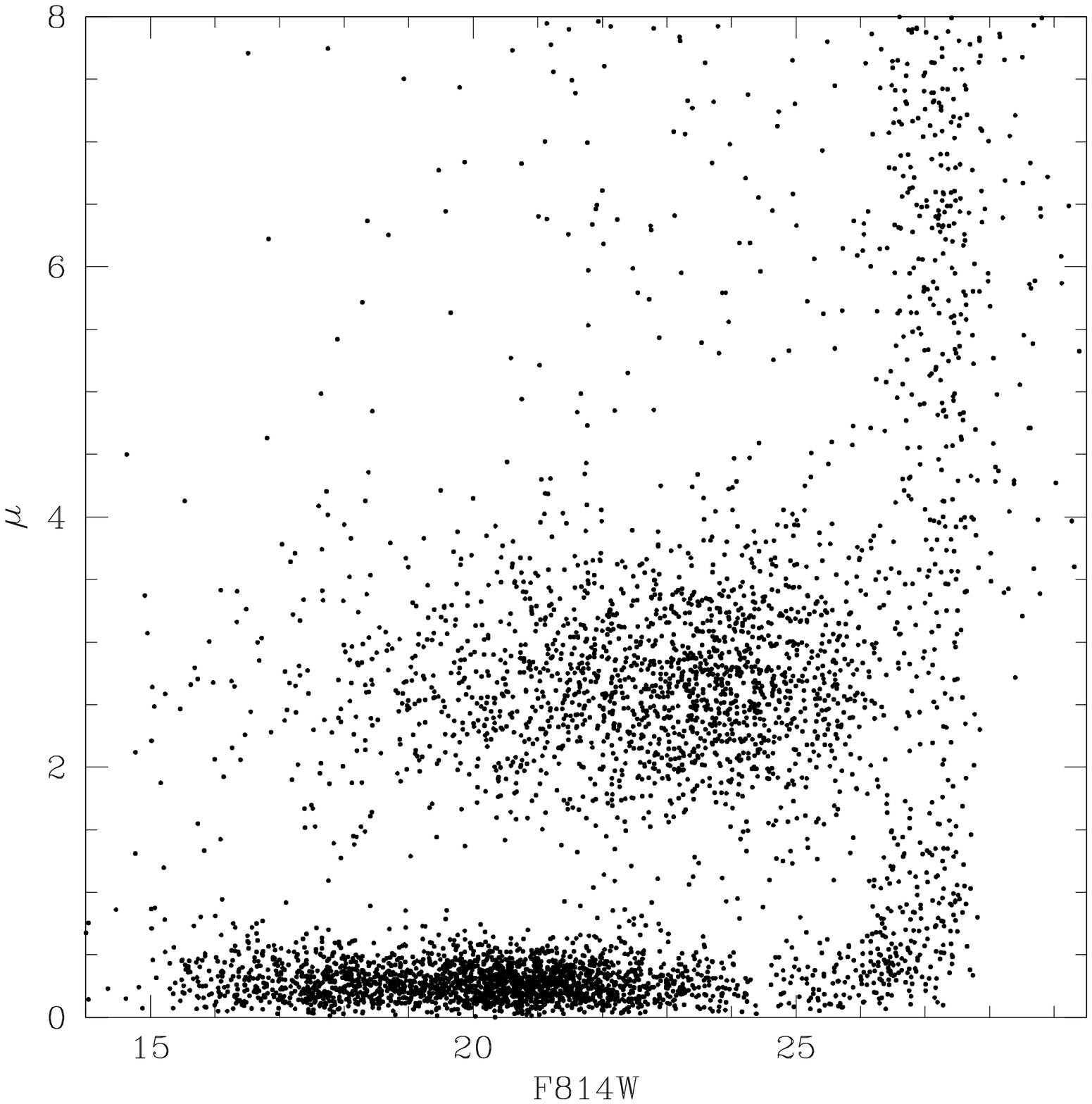}
\figcaption[IRO.ps]{ The proper motion displacements of all detected point sources that overlap
with WFPC2 data, as a function of F814W. We can see the excellent separation for
bright magnitudes, failing at F814W$ \sim 26.5$. Brighter than this magnitude, the stars separate
cleanly into cluster members ($\mu \sim 0$) and background stars ($\mu \sim 2.5$). Fainter than
this magnitude, most real stars in the ACS data are matched to noise peaks in the WFPC2 data, resulting
in large spurious displacements.
\label{IRO}}

\clearpage


\clearpage
\plotone{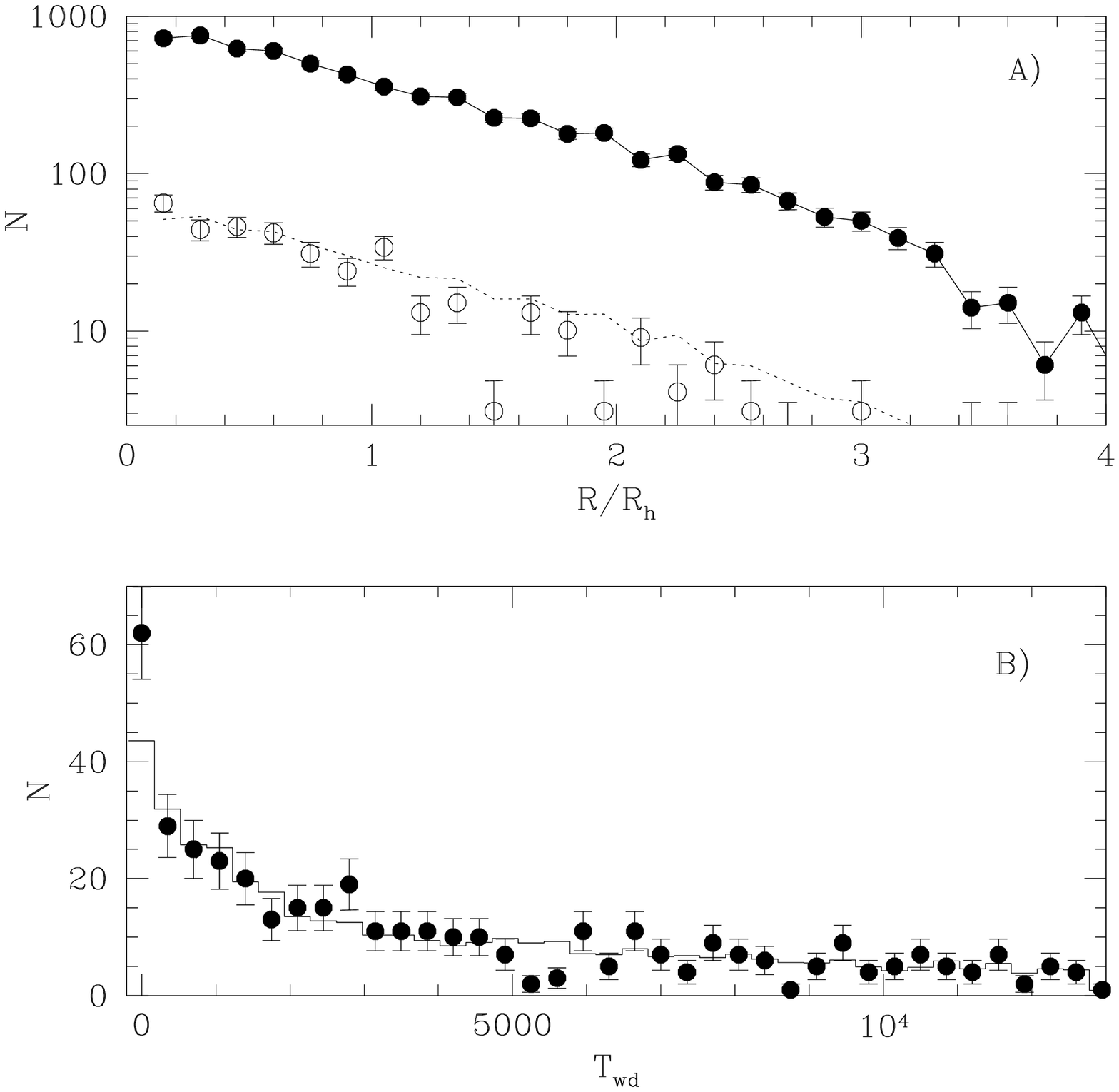}
\figcaption[2Pan.ps]{ A) The filled circles and solid line indicate the radial profile
of `bachelor' white dwarfs -- those which result from single star evolution and which
experience no perturbations during the course of their evolution. The open circles
represent the radial profile of `divorced' white dwarfs, which represent stars which
spent some fraction of their life in a binary but are now single. The dotted histogram
is the same as the solid histogram, but scaled by a factor 0.07. The divorced white dwarfs
follow the same radial profile as the bachelors. Radii are in units of the half-mass radius
of the cluster.
B) The filled circles show the distribution of divorced white dwarfs
with $\rm T_{wd}$ -- the time at which the white dwarf formed (i.e.$\rm T_{wd} = 0$ corresponds
to the birth of the cluster). The solid histogram represents the age distribution of the
bachelor white dwarfs, once again suitably scaled. There is clearly little difference in the
age distribution between the two populations. Even if the deviation at $\rm T_{wd}<200$~Myr is
real, it is too small to have any effect on our modeling.
\label{2Pan}}

\clearpage

\begin{deluxetable}{cllllll} 
\tablecolumns{7} 
\tablewidth{0pc} 
\tablecaption{NGC~6397 White Dwarf Distribution \label{CMDtable}}
\tablecomments{ Errors (statistical and systematic) in each bin are listed in parentheses.
} 
\tablehead{ 
\colhead{F814W}  & \multicolumn{6}{c}{F606W-F814W} \\
\cline{2-7} \\
 & \colhead{0.9--1.0} & \colhead{1.0--1.1} & \colhead{1.1--1.2} & \colhead{1.2--1.3} & \colhead{1.3--1.4} & \colhead{1.4--1.5}  \\
}
\startdata 
 Bright Bin   & 92 (13) &         &         &         &         &       \\
 26.00--26.25 & 36 (11) & 2 (2)   &         &         &         &       \\
 26.25--26.50 & 34 (13) & 52 (14) & 8 (5)   & 5  (3)  &         &       \\
 26.50--26.75 & 5 (5)   & 52 (13) & 77 (13) & 29 (7)  &         &       \\
 26.75--27.00 &         & 20 (10) & 65 (14) & 45 (10) & 14 (7)  & 3  (2) \\
 27.00--27.25 &         & 13 (8)  & 33 (10) & 83 (12) & 56 (11) & 17 (6)\\
 27.25--27.50 &         & 38 (16) & 72 (17) & 58 (14) & 62 (13) & 27 (10)\\
 27.50--27.75 &         & 30 (14) & 28 (13) & 28 (11) & 31 (9)  &        \\
 27.75--28.00 &         & 14 (10) & 8 (7)   & 4 (3)   & 4  (3)  &        \\
\enddata 
\end{deluxetable} 

\begin{deluxetable}{lcc}
\tablecolumns{3}
\tablewidth{0pc}
\tablecaption{Empirical NGC~6397 White Dwarf Cooling Sequence \label{EmpTab}}
\tablehead{
\colhead{F814W}  & \colhead{F606W-F814W} 
 & \colhead{$\Delta$(F606W-F814W) }  
 }
\startdata
 26.0 & 0.936 & 0.007 \\
 26.25 & 1.016 & 0.006 \\
 26.5  & 1.123 & 0.003 \\
 26.75 & 1.155 & 0.004 \\
 27.0 & 1.253 & 0.006 \\
 27.25 & 1.225 & 0.004 \\
 27.5 & 1.151 & 0.007 \\
\enddata
\end{deluxetable}

\begin{deluxetable}{lccccccccccccccccccc}
\rotate
\tablecolumns{20}
\tablewidth{0pc}
\tablecaption{F814W Photometric Scatter \label{DI}}
\tablecomments{Each bin corresponds to the difference between recovered magnitude and
input magnitude. Thus, a negative value implies that those stars were recovered as
being brighter than their true magnitudes.
}
\tablehead{
\colhead{F814W} & N$_{in}$ & N$_{out}$ & \multicolumn{17}{c}{$\Delta $F814W} \\
\cline{4-20} \\
& & & \colhead{-0.40} & \colhead{-0.35} & \colhead{-0.30} & \colhead{-0.25} & \colhead{-0.20} & \colhead{-0.15}
& \colhead{-0.10} & \colhead{-0.05} & \colhead{0.00} & \colhead{0.05} & \colhead{0.10} & \colhead{0.15} & \colhead{0.20}
& \colhead{0.25} & \colhead{0.30} & \colhead{0.35} & \colhead{0.40} \\
}
\startdata
  22.0 &    0 &    0 &    0 &    0 &    0 &    0 &    0 &    0 &    0 &    0 &    0 &    0 &    0 &    0 &    0 &    0 &    0 &    0 &    0  \\
  22.2 &    0 &    0 &    0 &    0 &    0 &    0 &    0 &    0 &    0 &    0 &    0 &    0 &    0 &    0 &    0 &    0 &    0 &    0 &    0  \\
  22.4 & 2750 & 2572 &    1 &    1 &    4 &    4 &   14 &   32 &   84 & 2333 &   83 &    5 &    1 &    1 &    2 &    1 &    0 &    0 &    0  \\
  22.6 & 2873 & 2684 &    3 &    0 &    3 &    4 &   11 &   14 &   55 & 2465 &  115 &    6 &    3 &    1 &    0 &    0 &    0 &    0 &    0  \\
  22.8 & 2917 & 2703 &    1 &    1 &    4 &    6 &   11 &   10 &   65 & 2487 &  101 &    5 &    3 &    0 &    0 &    1 &    0 &    0 &    0  \\
  23.0 & 2807 & 2610 &    0 &    1 &    4 &    4 &    6 &    8 &   31 & 2386 &  159 &    3 &    0 &    0 &    0 &    0 &    0 &    0 &    0  \\
  23.2 & 2781 & 2528 &    1 &    0 &    0 &    1 &    1 &    7 &   37 & 2239 &  230 &    8 &    3 &    0 &    0 &    0 &    0 &    0 &    0  \\
  23.4 & 2781 & 2525 &    0 &    0 &    3 &    2 &    2 &    4 &   21 & 2222 &  253 &   14 &    1 &    1 &    0 &    0 &    0 &    0 &    0  \\
  23.6 & 2866 & 2572 &    0 &    0 &    0 &    0 &    1 &    4 &   24 & 2205 &  316 &   18 &    3 &    0 &    0 &    0 &    0 &    0 &    0  \\
  23.8 & 2919 & 2631 &    0 &    0 &    0 &    1 &    1 &    3 &   12 & 2221 &  364 &   21 &    4 &    1 &    0 &    0 &    0 &    0 &    0  \\
  24.0 & 2805 & 2485 &    0 &    0 &    0 &    1 &    1 &    5 &   25 & 2035 &  390 &   20 &    5 &    0 &    0 &    0 &    0 &    0 &    1  \\
  24.2 & 2782 & 2446 &    0 &    0 &    0 &    0 &    1 &    4 &   38 & 1901 &  458 &   33 &    6 &    0 &    0 &    0 &    0 &    1 &    0  \\
  24.4 & 2774 & 2403 &    0 &    0 &    0 &    1 &    1 &    6 &   28 & 1839 &  477 &   34 &    9 &    4 &    0 &    0 &    0 &    0 &    0  \\
  24.6 & 2870 & 2460 &    1 &    0 &    1 &    2 &    5 &    4 &   32 & 1813 &  540 &   44 &   15 &    0 &    1 &    1 &    0 &    0 &    0  \\
  24.8 & 2915 & 2504 &    0 &    0 &    1 &    2 &    1 &    6 &   36 & 1763 &  628 &   51 &    9 &    3 &    1 &    0 &    0 &    0 &    0  \\
  25.0 & 2808 & 2349 &    0 &    0 &    0 &    1 &    1 &    5 &   46 & 1620 &  595 &   61 &   14 &    1 &    0 &    0 &    0 &    0 &    0  \\
  25.2 & 2784 & 2332 &    0 &    0 &    0 &    1 &    1 &    6 &   70 & 1436 &  707 &   81 &   20 &    3 &    1 &    0 &    0 &    0 &    0  \\
  25.4 & 2774 & 2249 &    0 &    0 &    0 &    0 &    2 &    8 &   55 & 1324 &  750 &   73 &   24 &    8 &    1 &    0 &    1 &    0 &    0  \\
  25.6 & 2879 & 2331 &    0 &    1 &    0 &    1 &    3 &   17 &   83 & 1325 &  783 &   84 &   19 &    9 &    2 &    0 &    0 &    0 &    0  \\
  25.8 & 2910 & 2322 &    0 &    1 &    0 &    3 &    3 &   21 &   74 & 1229 &  840 &   96 &   34 &   12 &    2 &    0 &    0 &    0 &    0  \\
  26.0 & 2806 & 2190 &    0 &    0 &    0 &    2 &    8 &   25 &   84 & 1087 &  815 &  108 &   36 &   13 &    5 &    1 &    0 &    1 &    0  \\
  26.2 & 2787 & 2112 &    1 &    2 &    3 &    3 &    7 &   21 &   89 &  988 &  814 &  130 &   35 &    8 &    1 &    0 &    0 &    0 &    0  \\
  26.4 & 2773 & 2042 &    0 &    1 &    3 &    2 &   10 &   20 &  116 &  884 &  785 &  139 &   43 &   19 &   12 &    1 &    1 &    0 &    0  \\
  26.6 & 2880 & 2075 &    1 &    0 &    4 &    4 &   12 &   36 &  169 &  803 &  792 &  169 &   59 &    8 &    7 &    2 &    0 &    0 &    0  \\
  26.8 & 2911 & 2041 &    1 &    0 &    5 &    7 &   21 &   39 &  163 &  762 &  720 &  232 &   53 &   21 &   10 &    2 &    0 &    0 &    0  \\
  27.0 & 2804 & 1888 &    0 &    0 &    3 &    9 &   21 &   46 &  185 &  629 &  631 &  230 &   79 &   33 &   14 &    3 &    1 &    0 &    0  \\
  27.2 & 2785 & 1751 &    0 &    2 &    3 &    9 &   27 &   55 &  202 &  530 &  515 &  254 &   92 &   30 &   18 &    4 &    4 &    2 &    0  \\
  27.4 & 2780 & 1665 &    0 &    1 &    5 &   10 &   21 &   77 &  216 &  418 &  448 &  271 &  112 &   47 &   20 &    7 &    5 &    1 &    0  \\
  27.6 & 2882 & 1636 &    1 &    3 &    7 &   15 &   35 &   95 &  233 &  370 &  381 &  263 &  128 &   60 &   26 &    9 &    4 &    2 &    1  \\
  27.8 & 2912 & 1541 &    1 &    7 &    7 &   22 &   38 &   99 &  209 &  311 &  315 &  244 &  154 &   65 &   41 &   12 &    2 &    6 &    2  \\
  28.0 & 2796 & 1350 &    1 &    2 &   12 &   16 &   54 &  106 &  172 &  234 &  251 &  186 &  143 &   77 &   46 &   22 &   12 &    4 &    4  \\
  28.2 & 2786 & 1178 &    2 &   12 &   15 &   27 &   58 &   97 &  157 &  170 &  191 &  148 &  119 &   85 &   37 &   21 &    5 &   10 &    4  \\
  28.4 & 2781 &  937 &    8 &   12 &   22 &   28 &   40 &   97 &  102 &  149 &  119 &  112 &   77 &   69 &   46 &   24 &    9 &    8 &    4  \\
  28.6 & 2882 &  718 &    7 &   14 &   23 &   31 &   53 &   75 &   91 &   85 &   73 &   80 &   63 &   41 &   21 &   15 &   10 &    5 &    6  \\
  28.8 & 2915 &  376 &    3 &   11 &   28 &   26 &   32 &   34 &   36 &   41 &   33 &   31 &   22 &   18 &   14 &   11 &    3 &    1 &    2  \\
  29.0 & 2790 &  222 &   10 &   14 &   20 &   14 &   28 &   17 &   19 &   14 &   21 &   10 &   13 &    4 &    6 &    0 &    3 &    0 &    1  \\
  29.2 & 2783 &  130 &    2 &   10 &   14 &   12 &   10 &   17 &    8 &   10 &    5 &    4 &    1 &    2 &    1 &    1 &    3 &    0 &    1  \\
  29.4 & 2783 &   53 &    3 &    5 &    1 &    4 &    7 &    2 &    0 &    1 &    3 &    0 &    3 &    2 &    0 &    0 &    1 &    1 &    0  \\
  29.6 & 2877 &   37 &    2 &    2 &    2 &    2 &    1 &    1 &    2 &    0 &    1 &    0 &    0 &    0 &    0 &    0 &    0 &    0 &    0  \\
\enddata
\end{deluxetable}

\begin{deluxetable}{lccccccccccccccccccc}
\rotate
\tablecolumns{20}
\tablewidth{0pc}
\tablecaption{F606W Photometric Scatter \label{DV}}
\tablecomments{Each bin corresponds to the difference between recovered magnitude and
input magnitude. Thus, a negative value implies that those stars were recovered as
being brighter than their true magnitudes.
}
\tablehead{
\colhead{F606W} & N$_{in}$ & N$_{out}$ & \multicolumn{17}{c}{$\Delta F606W$} \\
\cline{4-20} \\
& & & \colhead{-0.40} & \colhead{-0.35} & \colhead{-0.30} & \colhead{-0.25} & \colhead{-0.20} & \colhead{-0.15}
& \colhead{-0.10} & \colhead{-0.05} & \colhead{0.00} & \colhead{0.05} & \colhead{0.10} & \colhead{0.15} & \colhead{0.20}
& \colhead{0.25} & \colhead{0.30} & \colhead{0.35} & \colhead{0.40} \\
}
\startdata
  22.0 &    0 &    0 &    0 &    0 &    0 &    0 &    0 &    0 &    0 &    0 &    0 &    0 &    0 &    0 &    0 &    0 &    0 &    0 &    0 \\
  22.2 &  835 &  785 &    0 &    0 &    0 &    0 &    0 &    3 &   19 &  743 &   16 &    1 &    0 &    0 &    0 &    0 &    0 &    0 &    0 \\
  22.4 & 2533 & 2360 &    0 &    0 &    0 &    2 &    0 &    9 &   35 & 2238 &   72 &    2 &    1 &    0 &    0 &    0 &    0 &    0 &    0 \\
  22.6 & 2624 & 2440 &    0 &    0 &    1 &    1 &    0 &    4 &   29 & 2236 &  161 &    4 &    0 &    0 &    0 &    1 &    0 &    1 &    0 \\
  22.8 & 2599 & 2405 &    0 &    0 &    0 &    0 &    3 &    4 &   26 & 2081 &  277 &    1 &    1 &    1 &    0 &    0 &    0 &    0 &    0 \\
  23.0 & 2552 & 2366 &    0 &    2 &    1 &    0 &    1 &    3 &   11 & 1933 &  402 &    4 &    0 &    0 &    1 &    0 &    0 &    1 &    0 \\
  23.2 & 2523 & 2310 &    0 &    0 &    0 &    1 &    0 &    3 &   11 & 1772 &  520 &    1 &    0 &    0 &    0 &    0 &    0 &    0 &    0 \\
  23.4 & 2224 & 2007 &    0 &    0 &    0 &    0 &    1 &    0 &    2 & 1470 &  524 &    5 &    1 &    0 &    0 &    0 &    0 &    0 &    0 \\
  23.6 & 2268 & 2041 &    0 &    0 &    1 &    0 &    0 &    0 &    9 & 1488 &  538 &    1 &    0 &    0 &    0 &    0 &    0 &    0 &    0 \\
  23.8 & 2266 & 2037 &    0 &    0 &    0 &    0 &    0 &    0 &    7 & 1543 &  481 &    2 &    1 &    0 &    0 &    0 &    0 &    0 &    0 \\
  24.0 & 2275 & 2048 &    0 &    0 &    0 &    0 &    0 &    0 &    9 & 1536 &  486 &   14 &    0 &    0 &    0 &    0 &    0 &    0 &    0 \\
  24.2 & 2195 & 1948 &    0 &    0 &    0 &    0 &    0 &    0 &    3 & 1461 &  468 &   11 &    1 &    0 &    0 &    0 &    0 &    0 &    0 \\
  24.4 & 2280 & 1989 &    0 &    0 &    0 &    0 &    0 &    2 &   10 & 1500 &  460 &    9 &    3 &    1 &    0 &    0 &    0 &    0 &    0 \\
  24.6 & 2119 & 1873 &    0 &    0 &    0 &    0 &    0 &    1 &   25 & 1376 &  451 &   15 &    1 &    1 &    0 &    0 &    0 &    0 &    0 \\
  24.8 & 2153 & 1848 &    0 &    0 &    0 &    0 &    0 &    1 &    8 & 1355 &  457 &   14 &    5 &    0 &    2 &    0 &    0 &    0 &    0  \\
  25.0 & 2246 & 1921 &    0 &    0 &    0 &    0 &    1 &    2 &   14 & 1409 &  465 &   20 &    7 &    0 &    1 &    0 &    0 &    0 &    0  \\
  25.2 & 2258 & 1933 &    0 &    0 &    0 &    1 &    0 &    1 &   20 & 1415 &  464 &   28 &    2 &    1 &    0 &    0 &    0 &    0 &    0  \\
  25.4 & 2186 & 1844 &    0 &    1 &    0 &    0 &    0 &    3 &   20 & 1366 &  434 &   18 &    0 &    1 &    0 &    0 &    0 &    0 &    0  \\
  25.6 & 2197 & 1841 &    0 &    0 &    0 &    0 &    1 &    1 &   33 & 1341 &  428 &   26 &    4 &    1 &    3 &    0 &    0 &    0 &    0  \\
  25.8 & 2159 & 1806 &    0 &    0 &    0 &    2 &    2 &    8 &   42 & 1291 &  417 &   24 &   10 &    3 &    0 &    0 &    0 &    1 &    0  \\
  26.0 & 2006 & 1610 &    0 &    0 &    0 &    0 &    3 &    9 &   35 & 1157 &  362 &   29 &    6 &    3 &    3 &    1 &    0 &    0 &    0  \\
  26.2 & 2098 & 1702 &    0 &    0 &    1 &    1 &    2 &    9 &   49 & 1193 &  408 &   20 &   14 &    3 &    1 &    0 &    0 &    0 &    0  \\
  26.4 & 2098 & 1690 &    0 &    0 &    1 &    1 &    2 &   11 &   59 & 1107 &  451 &   32 &   11 &    3 &    5 &    0 &    1 &    0 &    1  \\
  26.6 & 2095 & 1660 &    0 &    1 &    0 &    1 &    3 &    9 &   69 & 1052 &  440 &   48 &   19 &    7 &    5 &    1 &    0 &    0 &    0  \\
  26.8 & 2028 & 1600 &    0 &    0 &    1 &    2 &    2 &   18 &   71 &  998 &  412 &   58 &   17 &   14 &    1 &    0 &    1 &    0 &    1  \\
  27.0 & 2070 & 1593 &    0 &    1 &    2 &    1 &    9 &   20 &  122 &  868 &  485 &   57 &   11 &    4 &    1 &    3 &    0 &    1 &    0  \\
  27.2 & 1985 & 1472 &    0 &    1 &    1 &    1 &    8 &   18 &  124 &  739 &  436 &   80 &   26 &   14 &    6 &    4 &    0 &    1 &    3  \\
  27.4 & 2144 & 1588 &    1 &    1 &    0 &    4 &   13 &   29 &  156 &  728 &  480 &  100 &   36 &   18 &   11 &    2 &    1 &    1 &    1  \\
  27.6 & 2228 & 1609 &    0 &    1 &    3 &    6 &    9 &   48 &  188 &  675 &  491 &  114 &   36 &   10 &    7 &    5 &    5 &    2 &    2  \\
  27.8 & 2266 & 1602 &    1 &    2 &    5 &    7 &   18 &   41 &  233 &  571 &  472 &  151 &   48 &   17 &   10 &    5 &    7 &    3 &    0  \\
  28.0 & 2148 & 1518 &    1 &    2 &    7 &   10 &   24 &   73 &  208 &  502 &  404 &  174 &   56 &   23 &   11 &    6 &    4 &    2 &    2  \\
  28.2 & 2223 & 1469 &    1 &    4 &    4 &   11 &   27 &   92 &  208 &  381 &  342 &  197 &   94 &   39 &   26 &    9 &    6 &    2 &    1  \\
  28.4 & 2199 & 1397 &    2 &    0 &    6 &   18 &   27 &   89 &  201 &  312 &  330 &  194 &  107 &   47 &   20 &    9 &    6 &    6 &    6  \\
  28.6 & 2091 & 1236 &    6 &    3 &   10 &   14 &   39 &  100 &  171 &  241 &  240 &  174 &  104 &   46 &   33 &   19 &    5 &    5 &    3  \\
  28.8 & 2207 & 1310 &    2 &    6 &   13 &   26 &   56 &  114 &  182 &  226 &  220 &  156 &  101 &   74 &   38 &   29 &   16 &   10 &    8  \\
  29.0 & 2238 & 1247 &    6 &    6 &   21 &   33 &   64 &  111 &  134 &  195 &  179 &  158 &  102 &   73 &   52 &   30 &   24 &   12 &    7  \\
  29.2 & 2273 & 1202 &   10 &   13 &   22 &   45 &   70 &  109 &  147 &  150 &  153 &  110 &  106 &   63 &   61 &   34 &   29 &   20 &    4  \\
  29.4 & 2184 & 1088 &   12 &   20 &   34 &   51 &   92 &   84 &   98 &  115 &   90 &  109 &   91 &   62 &   52 &   39 &   37 &   26 &   14  \\
  29.6 & 2184 &  994 &   15 &   22 &   33 &   46 &   62 &   72 &  105 &   89 &  103 &   71 &   60 &   64 &   48 &   35 &   26 &   26 &   11  \\
\enddata
\end{deluxetable}

\end{document}